\documentclass[journal=jpccck,manuscript=article]{achemso}

\usepackage[version=3]{mhchem}

\usepackage{tabularx}
\usepackage{multirow}

\newcommand{\superscript}[1]{\ensuremath{^{\textrm{#1}}}}
 \newcommand{\subscript}[1]{\ensuremath{_{\textrm{#1}}}}

\author{Conn O'Rourke}
\email{ucapcor@ucl.ac.uk}
\affiliation[LCN]
{London Centre for Nanotechnology, 17-19 Gordon St, London, WC1H 0AH}
\alsoaffiliation[UCL]
{Department of Physics \& Astronomy, University College London, Gower St,
London, WC1E 6BT}
\alsoaffiliation[UCL Satellite, MANA]
{UCL Satellite, International Centre for Materials Nanoarchitectonics (MANA), National Institute for Materials Science (NIMS), 1-1 Namiki, Tsukuba, Ibaraki 305-0044, Japan}
\author{David R. Bowler}
\affiliation[LCN]
{London Centre for Nanotechnology, 17-19 Gordon St, London, WC1H 0AH}
\alsoaffiliation[UCL]
{Department of Physics \& Astronomy, University College London, Gower St,
London, WC1E 6BT}
\alsoaffiliation[UCL Satellite, MANA]
{UCL Satellite, International Centre for Materials Nanoarchitectonics (MANA), National Institute for Materials Science (NIMS), 1-1 Namiki, Tsukuba, Ibaraki 305-0044, Japan}
\title[]{ Intrinsic Oxygen Vacancy and Extrinsic Aluminium Dopant Interplay: A Route to the Restoration of Defective TiO\subscript{2}}

\begin{document}

   \begin{abstract}

Density functional theory (DFT) and DFT corrected for 
on-site Coulomb interactions (DFT+U) calculations are presented 
on Aluminium doping in bulk TiO\subscript{2} and the anatase (101) surface.
Particular attention is paid to the 
mobility of oxygen vacancies throughout the doped TiO\subscript{2}
lattice, as a means by which charge compensation of trivalent dopants can occur. The effect that Al doping of TiO\subscript{2} electrodes has in 
dye sensitised solar cells is explained as a result of this
mobility and charge compensation.
Substitutional defects in which one Al\superscript{3+} replaces one
Ti\superscript{4+} are found to introduce valence band holes, while
intrinsic oxygen vacancies are found to introduce states in the band-gap.
Coupling two of these substitutional defects with an oxygen vacancy results in exothermic defect formation which maintain charge neutrality. 
Nudged elastic
band calculations have been performed to investigate the formation of these clustered
defects in the (101) surface by oxygen vacancy diffusion, with the resulting potential energy surface suggesting energetic gains
with small diffusion barriers. 
Efficiency increases observed in dye sensitised solar cells as a result 
of aluminium doping of TiO\subscript{2} electrodes are investigated by adsorbing the 
tetrahydroquinoline C2-1 chromophore on the defective surfaces. 
Adsorption on the clustered extrinsic Al\superscript{3+} and intrinsic oxygen vacancy defects are found to behave as if adsorbed on a clean surface, with vacancy states not present, while adsorption on the oxygen vacancy
results in a down shift of the dye localised states within the band-gap
and defect states being present below the conduction band edge. 
Aluminium doping therefore acts as a benign dopant for 'cleaning' TiO\subscript{2} through oxygen vacancy diffusion.

    \end{abstract}

\section{Introduction}

Titanium dioxide (TiO\subscript{2}) has a wide variety of technological uses 
to which the considerable scientific interest in its surface properties can
 be attributed.
For example TiO\subscript{2} is used
in photocatalysis\cite{Fujishima-Photocatalysis,Asahi13072001} and in
dye sensitised solar cells (DSSCs)\cite{Gratzel}. DSSCs have been receiving
widespread attention recently as a possible
a clean, cost-effective renewable energy source.

Crystal defects can have a significant role in defining the properties of 
TiO\subscript{2}, and therefore the electrodes used in DSSCs. 
Actively doping TiO\subscript{2} with nitrogen, for example,
 is known to lower the photo-excitation threshold  
in anatase TiO\subscript{2} \cite{Valentin2006,Valentin2005}, an extrinsic 
defect which has been put to use in 
photo-catalysis\cite{Asahi13072001}. Aluminium dopants can be introduced 
in TiO\subscript{2} by
inclusion of aluminium butoxide during the hydrolysis of  titanium iso-propoxide (TTIP) \cite{Ko-AL-DSSC} to form TiO\subscript{2}.
DSSC electrodes including aluminium dopants produced in this manner have
 been shown to decrease the concentration of Ti\superscript{3+} defects 
resulting from oxygen 
vacancies, improve the open circuit voltage (V\subscript{OC}) and thereby the DSSC efficiency \cite{Ko-AL-DSSC}, however 
the mechanism has not been fully understood.  
Intrinsic defects, such as oxygen vacancies, also have
a important role in the chemical reactivity of TiO\subscript{2}
surfaces. An example is dissociation of water at vacancy sites
on rutile (110) \cite{Henderson1998203}.

Oxygen vacancies introduce localised band-gap states in TiO\subscript{2} resulting
in the formation of Ti\superscript{3+} ions
 \cite{Nakamura2000,Lin-OVac} which can trap injected 
electrons, and act as recombination centres \cite{Weidmann1998153}.
Oxygen plasma treatments of TiO\subscript{2} 
electrodes, which reduce the number of oxygen vacancies, have shown a marked 
increase in the short-circuit current of DSSCs \cite{Kim-OvacPLASMA}, 
which indeed suggests that 
vacancies have a negative effect on DSSC performance. An increase in recombination sites in a DSSC will lead to interception of injected electrons by either the 
redox couple in solution, or by back transfer to dyes. Similarly this increase 
in recombination can cause a down shift in the quasi-Fermi energy of electrons in the conduction band, and a subsequent reduction in the short circuit voltage
V\subscript{OC}.

Previous theoretical studies carried out on aluminium doped TiO\subscript{2} have 
examined the stability of bulk defects in both rutile and anatase. In the case of 
anatase, clustering of defects in which two Al dopants combined with oxygen vacancies was found to produce the most stable defect type \cite{Shirley-AL-DOP,Bredow-AL-DOP}.
However the migration of aluminium interstitials throughout the bulk anatase 
crystal was found impossible at industrial temperatures, due to large transition 
barriers \cite{Shirley-AL-DOP}.
 Oxygen vacancies are known to preferentially occupy sub-surface 
sites in anatase (101) \cite{Selloni-VAC1,Selloni-VAC2}, with diffusion
 barriers that can be overcome at typical 
annealing temperatures. Diffusion of these oxygen vacancies suggests another mechanism
by which stable clustering of intrinsic and extrinsic defects may occur in aluminium 
doped TiO\subscript{2}.

The object of this work is to examine the effect of aluminium doping on the majority 
anatase (101) surface\cite{Zhang-ANATASE}, and to understand the observed increase 
in DSSC efficiency which
 results. To this end density functional theory (DFT) calculations have
been carried out to characterise the doped (101) surface, with a particular focus on the
interplay between intrinsic oxygen vacancy defects and the extrinsic aluminium dopants. 
Defect stabilities have been calculated for  
aluminium defects with and without the presence of oxygen 
vacancies, in both the bulk and the (101) surface. 

Nudged elastic band (NEB) calculations have also been performed to establish diffusion barriers for oxygen vacancies in the presence of aluminium dopants, and illustrate the possibility of the intrinsic
extrinsic defect clustering. Finally the effect of these defects on the adsorption of a typical DSSC dye is examined.


\section{Computational Detail}

All calculations have been performed using the plane wave DFT 
code VASP \cite{Vasp}. Exchange and correlation effects were 
approximated by the 
generalised gradient approximation of Perdew and Wang \cite{PW91} with 
core electron wavefunctions treated within the projector augmented wave method 
\cite{PAW}. For both bulk and surface calculations a plane wave cut-off
of 800 eV has been used, which we have tested and give total energies converged to within 0.06 eV of values at 700 eV for the bulk.

\begin{table}
\begin{center}
\begin{tabular}{c c c c}
\hline
\hline
& Expt. \cite{anatase_exp} & Ref (US PP's) \cite{Conn-Thio}& This Work\\
& (\AA) &(\AA)&(\AA)\\
\hline
Lattice Parameters &&&\\
a & 3.782& 3.817 (+0.9\%)& 3.804 (+ 0.6\%)\\
c & 9.502& 9.710 (+2.2\%)& 9.698 (+ 2.1\%)\\
Bond Lengths&&&\\
Equatorial & 1.932& 1.954 (+1.1\%)& 1.947 (+ 0.8\%)\\
Apical & 1.979& 2.011 (+1.6\%)& 2.005 (+ 1.3\%)\\
\hline
\hline
\end{tabular}
\end{center}
\caption{Calculated and experimental lattice parameters and bond lengths of bulk anatase
TiO\subscript{2}}
\label{Anatase_bulk}
\end{table}

Calculated lattice parameters for the bulk anatase TiO\subscript{2}
without defects can be seen in Fig. \ref{Anatase_bulk}.
Good agreement with experimental data and
the results from the previous work can be seen, and we have
used these bulk
lattice parameters throughout the current investigation.
 
Periodic images of the defects will interact with one another, and in order to gauge 
the extent of this interaction in the bulk case we have taken one defect type (A2; see
Fig. \ref{Defects}) and performed calculations for varying supercell dimensions.
Supercells containing one defect with sizes of 2, 3, 4 and 5 unit cells have been
examined.
Supercell extension was along one minor lattice vector (a=3.817 \AA) while
the other two vectors were kept constant.
A Monkhorst-Pack k-point mesh has been 
utilised throughout these calculations with dimensions of $6 \times 6 \times 3$ for 
the 
smallest supercell, varying to $3 \times 6 \times 3$ for the largest supercell and 
provided energies converged to within $1\times10^
{-5}$ eV. 
 Relaxations were performed using the conjugate
gradients method, and considered finished when forces on ions were less than
0.03 eV/\AA. 

\begin{figure}
\begin{center}
   \begin{minipage}[t]{0.45\textwidth}
 \includegraphics[trim = 0mm 0mm 0mm 0mm, clip, width=1.\textwidth]{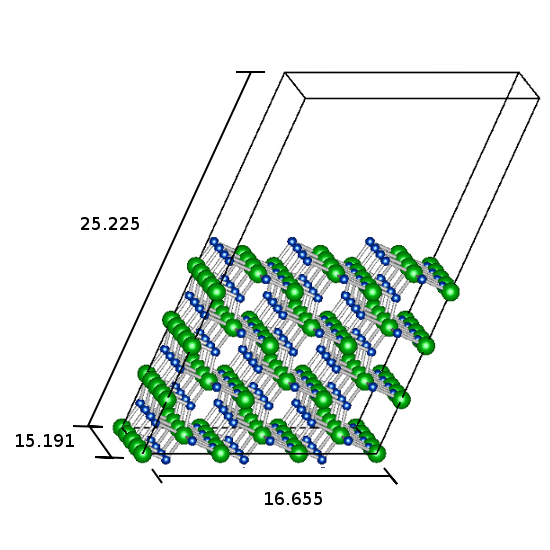}
  \end{minipage}
\caption{Anatase (101) surface: 4 layer slab used in defect calculations.
 Titanium atoms are represented in green, oxygen atoms in blue. 
The bounding box shows the unit cell with the cell dimensions in \AA.}
 \label{Ana-layer}
\end{center}
\end{figure}

For surface calculations a four layer triclinic unit cell containing
 288 atoms, with 
cell dimensions illustrated in Fig. \ref{Ana-layer}, was the clean starting point.
In order to prevent spurious interactions between periodic slabs 
it was ensured that a vacuum layer of at least 10 \AA\ 
 separated images in the (101) direction for all calculations.
A Monkhorst-pack grid of 
$2 \times 2 \times 1$ was utilised throughout surface calculations, and increased to 
$4 \times 4 \times 1$ for density of states calculations.
Fixing the bottom layer to the relaxed bulk position, geometry relaxations 
were again
performed with the conjugate gradient method until ionic forces were less than 0.03
eV/\AA. Calculations on  chromophore adsorption were performed 
following previous work \cite{Conn-Thio}, and considered fully relaxed when forces on ions were less than 0.03 eV/\AA.
In the calculation of defect stability isolated molecule calculations have been performed, and in each case the cell dimensions are the same as the substrate to which they will be compared.

Results reported for the A2 and 
A3 defects have been obtained with spin unpolarised calculations. 
In the case of the A1 defect, the single substitution of an
Al dopant leaves one unpaired electron so spin polarised 
calculations are performed throughout.

Semi-local functionals such as PW91 are known to incorrectly describe the Ti\superscript{3+} states resulting from an oxygen vacancy in
TiO\subscript{2}, due largely to self-interaction errors and the band gap underestimation\cite{GGA-Ovac}. A similar failure has been reported in the case of single
aluminium dopants for TiO\subscript{2} in the Rutile phase\cite{iwaszuk}.
While employing hybrid functionals has been shown to more accurately 
describe these localised states in defective TiO\subscript{2}, in plane wave codes such as VASP the use of hybrid functionals introduces an 
extra order of magnitude of computational time, which is a significant 
burden on the computational scientist. 

We do not feel that the extra computational effort is justified here. As
such as a step to rectify the limitations of GGA in the case of single Al substitutional defects and isolated
oxygen vacancies, we have
also performed GGA corrected for on-site Coulomb interactions (GGA+\emph{U}) \cite{DFT+U,DFT+U2}, employing the GGA+U correction in the form
of Dudarev\cite{DFT+U}. Hybrid functionals themselves 
are not fully characterised, as the fraction of exchange varies 
between functionals (which makes them, at least in principle, 
as empirical as DFT+U) and we feel that GGA+\emph{U} is the more
pragmatic approach given the size of the systems under study.

While recent developments have suggested a 
route to the calculation of U self-consistently\cite{DFTwannier}, there is insufficient 
data on this approach to trust it absolutely. 
Therefore we must fit the value of U used to the problem under consideration, and this is one of the standard approaches to its use.
As it is a relatively simple model, a single value of U will not
 fit different circumstances (and indeed a self-consistent U would 
give different values in different environments). 
Therefore in the case of the oxygen vacancy U with a value of 3eV has been
applied to the Ti 3d orbitals, which has been shown to
qualitatively agree with the B3LYP hybrid functional in the
case of oxygen vacancies in Anatase TiO\subscript{2}\cite{OvacU}.
A range of values of \emph{U} are applied to the O 2p states in the case
of a single Al\superscript{3+} for Ti\superscript{4+} substitution in the bulk, with
a value of 6eV used for surface calculations.

\section{Bulk Defects}

\begin{figure}
\begin{center}
\begin{tabular}{c c c  c}
  \begin{minipage}[t]{0.14\textwidth}
  \includegraphics[trim = 30mm 0mm 85mm 0mm, clip, width=1.\textwidth]{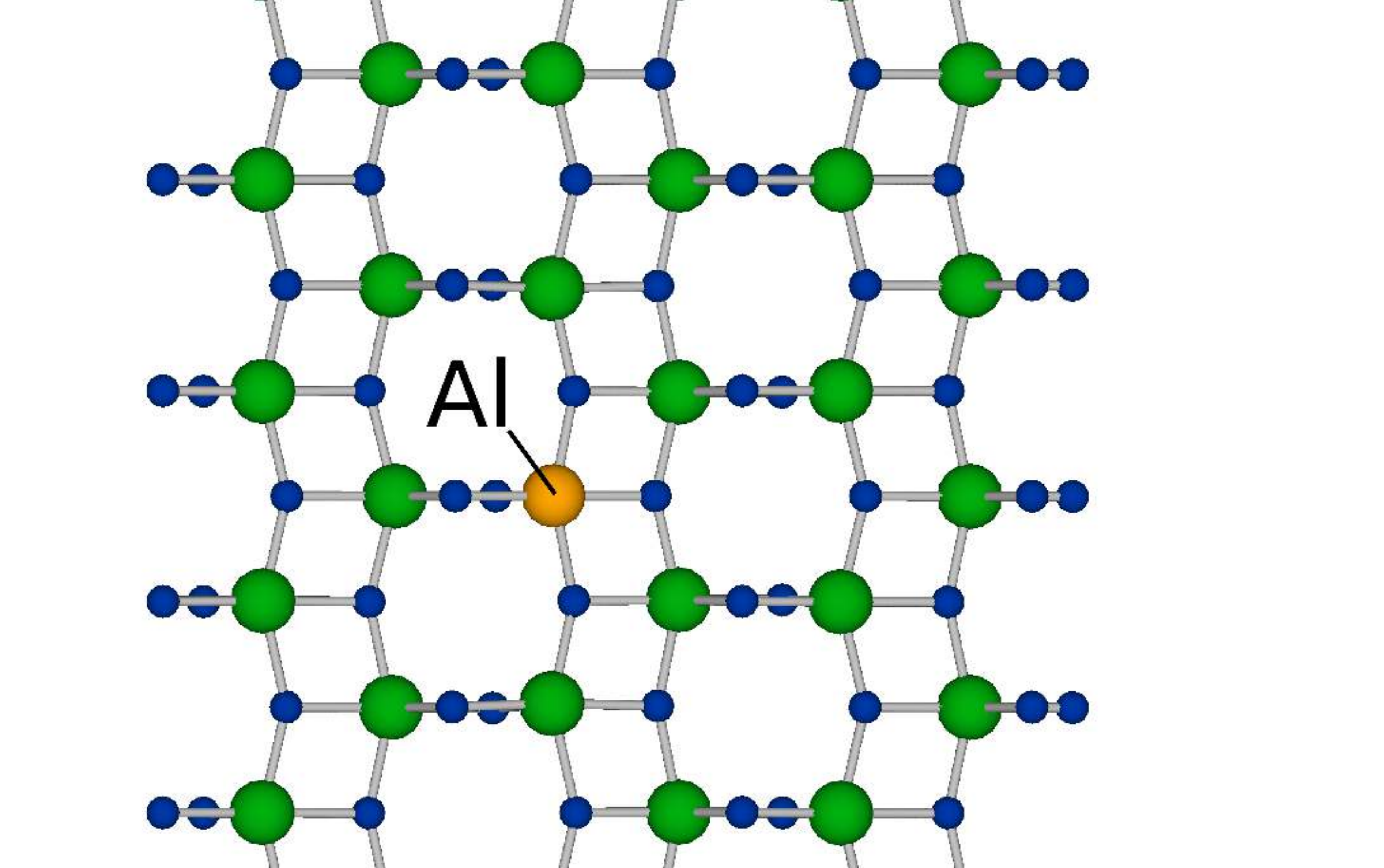}
  \end{minipage}
  &
  \begin{minipage}[t]{0.14\textwidth}
  \includegraphics[trim = 30mm 0mm 85mm 0mm, clip, width=1.\textwidth]{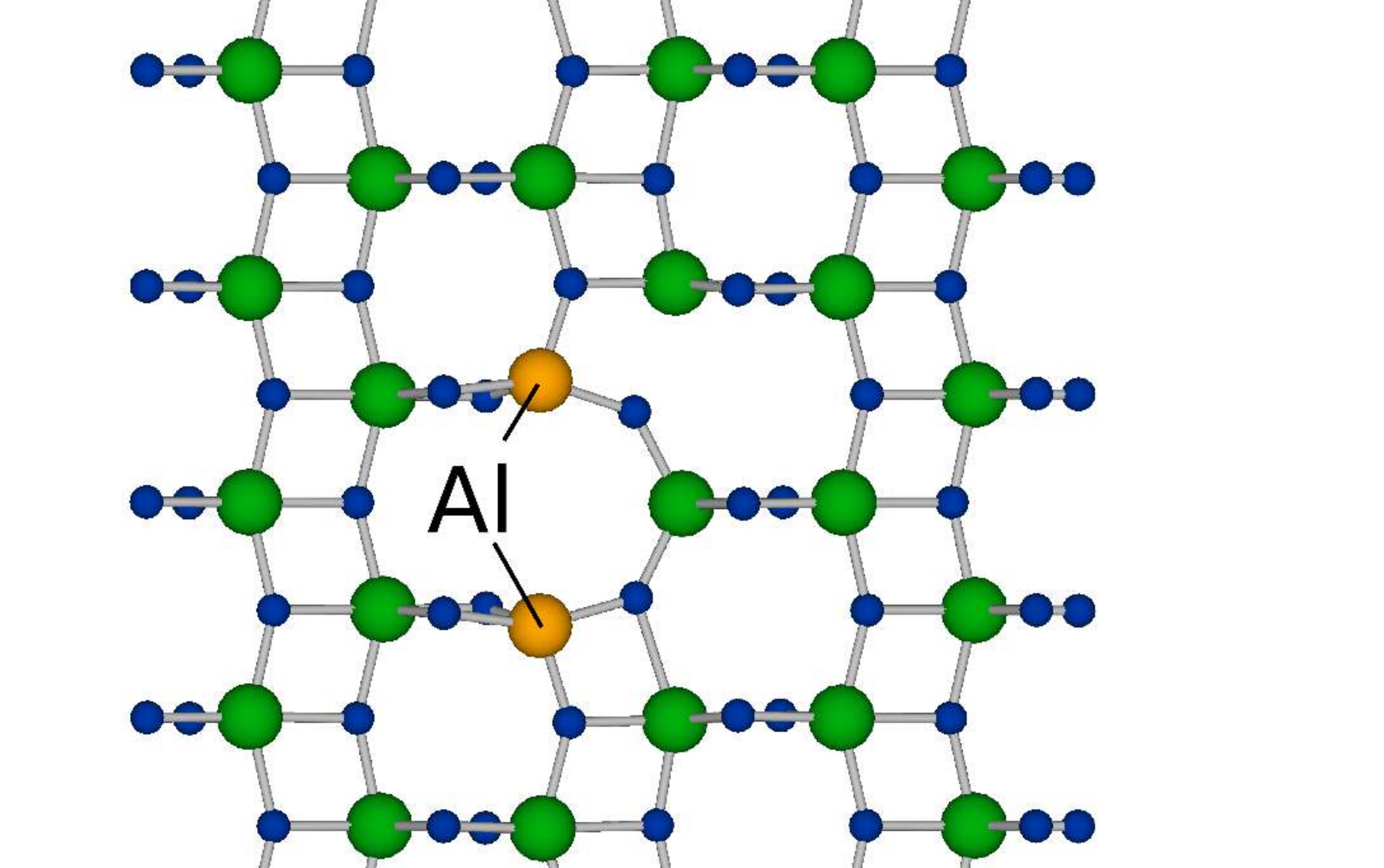}
  \end{minipage}
  &
  \begin{minipage}[t]{0.14\textwidth}
  \includegraphics[trim = 30mm 0mm 85mm 0mm, clip, width=1.\textwidth]{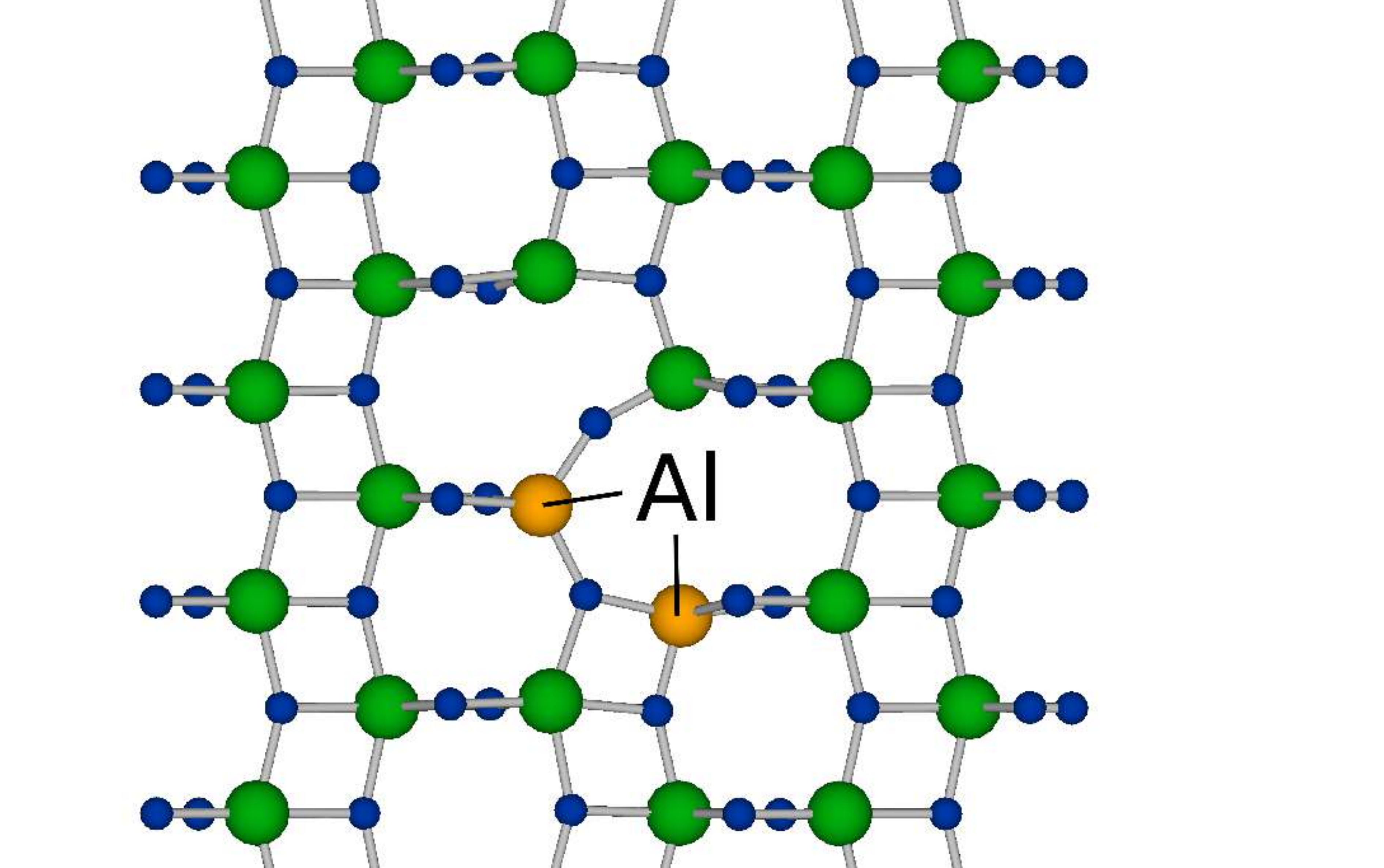}
  \end{minipage}
  &
  \begin{minipage}[t]{0.14\textwidth}
  \includegraphics[trim = 30mm 0mm 5mm 0mm, clip, width=1.\textwidth]{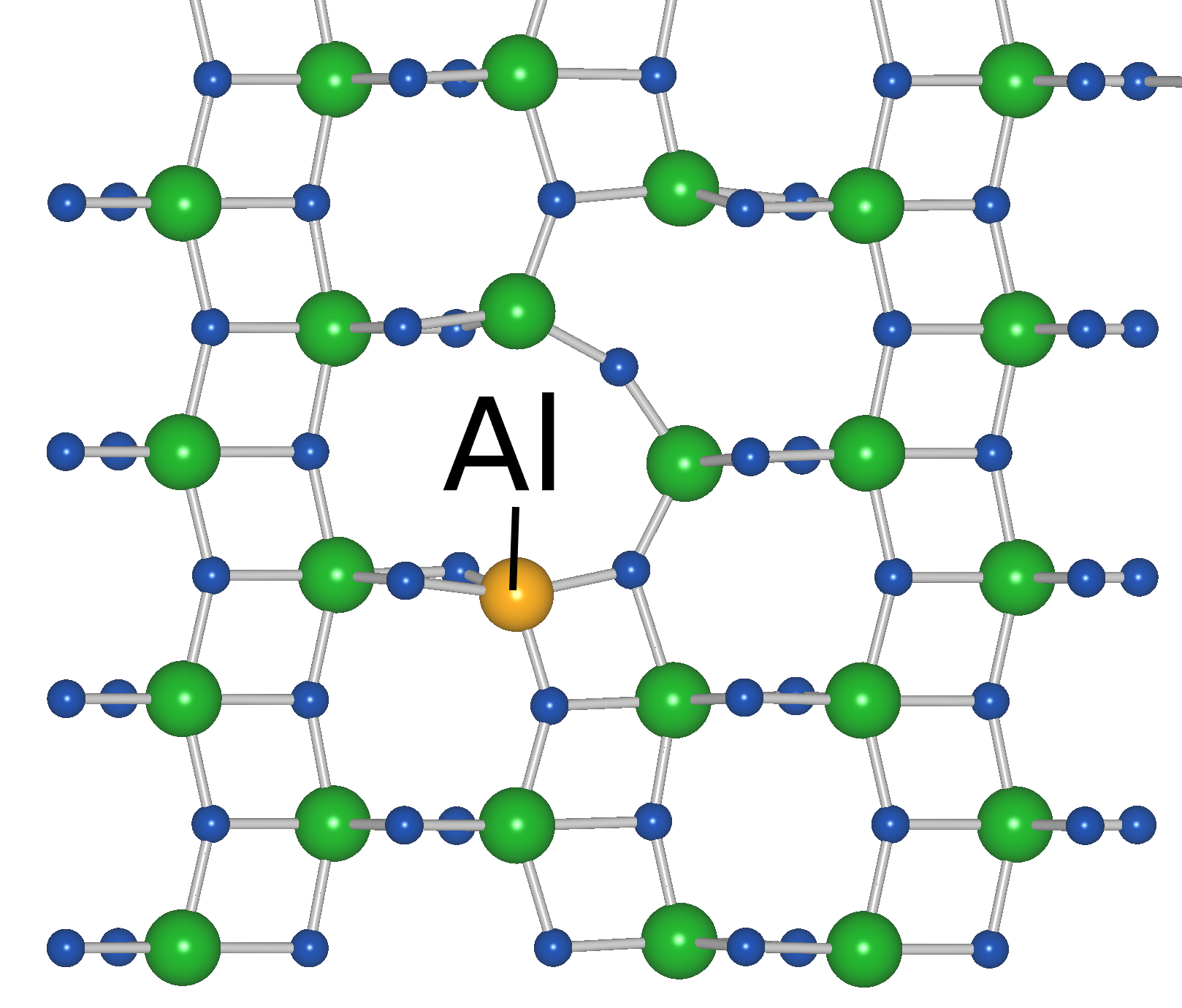}
  \end{minipage}
\\
  \textbf{A1}&\textbf{A2}&\textbf{A3}&\textbf{A4}\\
\end{tabular}
\caption{Aluminium doped anatase defect structues. Titanium atoms are represented in green, oxygen in blue, aluminium in orange. For clarity the defects are shown 
in a single (101) layer.}
 \label{Defects}
\end{center}
\end{figure}

Previous studies have suggested that interstitial defects are
 relatively less stable
than substitutional doping \cite{Shirley-AL-DOP}, so here we focus only on these
substitutional defect types. 
This is consistent with experimental evidence for the aluminium doping of 
rutile suggesting that Al will substitutionally replace 
Ti atoms\cite{rut_subst,rut_subst2}.
Four different defects are considered in the bulk
(illustrated in Fig. \ref{Defects}); direct substitution of a titanium atom with an
 aluminium atom (A1), two substitutional defects combined with an oxygen vacancy 
(A2 \& A3) and a single substitutional defect combined with an oxygen vacancy (A4).

Alminium dopants may also be introduced in TiO\subscript{2} during the growth from the combustion of TiCl\subscript{4} by inclusion of AlCl\subscript{3}. 
As in the work of Shirley et al \cite{Shirley-AL-DOP}, we use as a measure of defect stability the energy of reaction for the following equivalent reaction: 

\begin{align}
\left(x+z\right)\mbox{TiO}_{2}+z\mbox{AlCl}_{3}+\frac{z}{2}\mbox{Cl}_{2} &\rightleftharpoons \nonumber\\ 
\mbox{Ti}_{x}\mbox{O}_{y}\mbox{Al}_{z}+ z\mbox{TiCl}_{4}&+\left(x+z-\frac{y}{2}\right)\mbox{O}_{2} 
\end{align}

such that
 
\begin{align}
\Delta \mbox{E}_{0K}= &\mbox{E}_{b}\left(\mbox{Ti}_{x}\mbox{O}_{y}\mbox{Al}_{z}\right)+ \mbox{E}_{b}\left(z\mbox{TiCl}_{4}\right)\nonumber\\
&+\mbox{E}_{b}\left(\left(x+z-\frac{y}{2}\right)\mbox{O}_{2}\right)-\mbox{E}_{b}\left(\frac{z}{2}\mbox{Cl}_{2}\right) \nonumber\\
&-\mbox{E}_{b}(\left(x+z\right)\mbox{TiO}_{2})-\mbox{E}_{b}\left(z\mbox{AlCl}_{3}\right)
\end{align}
 
where $\Delta \mbox{E}_{0K}$ is the energy of reaction at 0K, 
and E\subscript{b} is 
binding energy as given by our DFT calculations. Zero point energies
have been neglected. 

As a first step we perform calculations on the A2 defect in order to
evaluate the typical defect-defect interaction. Calculated bulk
energies of reaction for the A2 defects in varying supercell sizes
 are exhibited in Table \ref{Bulk-Defect-A2}. 
Reaction energies suggest that
defect-defect interaction is fairly short ranged, with
energies at a defect separation along the varying vector
 of around 7\AA\ converged
to within 0.06 eV of that at a distance of around 11\AA.
In order to minimize the defect-defect interaction bulk calculations
proceed with supercells of dimenisons $4 \times 2 \times 1$ unit cells, 
containing 96 atoms in the clean supercell,
 giving a defect separation of greater than 7.5\AA\ in all directions.
Experimentally the atomic decomposition of the doped powder made up of
3.3\% aluminium\cite{Ko-AL-DSSC}, while cell dimensions
$4 \times 2 \times 1$ give us 2.1\% for A2 \& A3 defects.

\begin{table}[t]
\begin{center}
\begin{tabular}{c c c }
\hline
\hline
Defect Type &  A2 &\\
85            & (eV)&\\
\hline
Supercell &  &Defect Separation\\
Dimension &  & (\AA)\\
\hline
$2 \times 2 \times 1$    & -0.051  & 3.81\\
$3 \times 2 \times 1$    & -0.280  & 7.62\\
$4 \times 2 \times 1$    & -0.342  & 11.44\\
$5 \times 2 \times 1$    & -0.368  & 14.77\\
\hline
\hline
\end{tabular}
\caption{Calculated $\Delta\mbox{E}_{0K}$ for bulk aluminium defect typei A2 with varying supercell size.
Dimensions are given as multiples of unit cell vectors along the two minor and
one major axis respectively. }
\label{Bulk-Defect-A2}
\end{center}
\end{table}

The calculated defect stability for each of the four defects examined can be seen in 
Table \ref{Bulk-Defect2}, and in the following subsections we discuss each of these bulk
defects in turn.

\subsection{A1 defect}

\begin{figure}[t]
\begin{center}
 \begin{tabular}{c c}
\begin{minipage}[c]{0.4\textwidth}
 \centering
  \begin{minipage}[c]{0.35\textwidth}
  \includegraphics[angle=0,origin=c,trim = 0mm 0mm 0mm 10mm, clip, width=1.\textwidth]{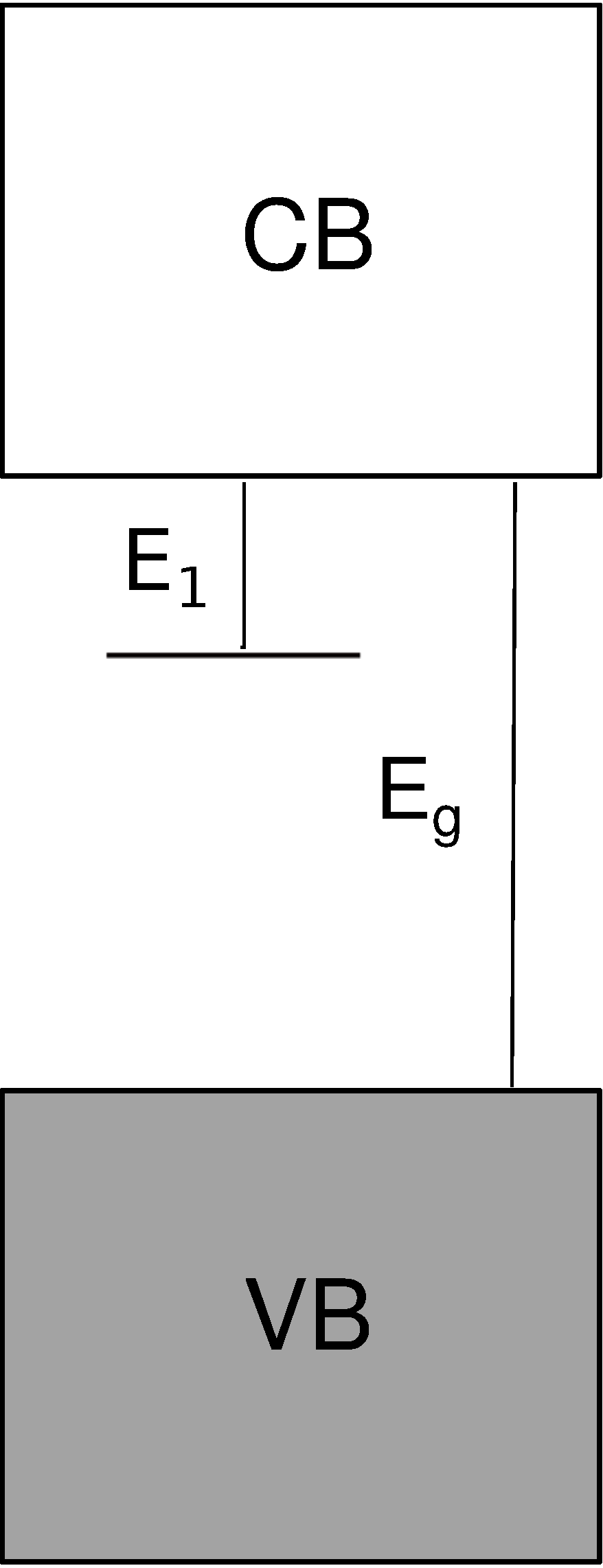}
  \end{minipage}
  \end{minipage}
&
 \begin{minipage}[c]{0.6\textwidth}
\centering
\begin{minipage}[c]{0.9\textwidth}
\centering
 \begin{tabular}{c c c c}
\hline
\hline
 &E\subscript{1}&E\subscript{g}&$\Delta\mbox{E}_{0K}$\\
\emph{U}(eV) &\multicolumn{3}{c}{(eV)}\\
\hline
0&0.00&2.17&0.293\\
1&N/A&N/A&N/A\\
2&1.78&2.17&0.232\\
3&1.50&2.19&0.093\\
4&1.21&2.21&-0.076\\
5&0.92&2.23&-0.266\\
6&0.64&2.25&-0.472\\
7&0.39&2.26&-0.689\\
\hline
\hline
\end{tabular}
 \end{minipage}
\end{minipage}
\end{tabular}
\caption{\textbf{A1 Bulk Defect:} \emph{U} dependence on band gap (E\subscript{g}), oxygen hole state (E\subscript{1}) and defect formation energy ($\Delta\mbox{E}_{0K}$).
}
 \label{A1_U}
\end{center}
\end{figure}

Substitution of an Al\superscript{3+} ion for a Ti\superscript{4+} ion
will result in one less electron in the system, and an oxygen
hole being formed. 
Polaronic in nature, this resulting
O\superscript{-} state is poorly described by GGA and
we have examined the defect stability and hole characteristics with
varying values of on-site Coulomb interaction (\emph{U}) Fig. \ref{A1_U}.



\begin{figure}[t]
\begin{center}
 \begin{tabular}{c}
\footnotesize
\hspace{-15mm}
\begin{minipage}[t]{1.0\textwidth}
\begin{center}
\hspace{-5mm} \begin{tabular}{c c c c c c c c c c c }
\hline
\hline
&\multicolumn{9}{c}{Bond (\AA)}\\
& Al-O1 & Al-O2 & Al-O3 & Al-O4 & Al-O5 & Al-O7 & O5-Ti2 & O5-Ti3& O7-Ti4 & O7-Ti5 \\
\hline
Method&&&&&&&&&\\
\hline
GGA &1.91&1.90&1.93&1.93&1.91&1.91&2.04&1.88&2.02&1.89\\
GGA+\emph{U}(6 eV.) &1.90&1.85&1.93&1.91&\textbf{1.95}&1.90&\textbf{2.29}&\textbf{2.07}&2.01&1.90\\
\hline
\hline
\end{tabular}
\end{center}
\end{minipage}
\normalsize
\\
  \begin{tabular}{c c c}

   \begin{minipage}[c]{0.27\textwidth}
      \includegraphics[trim = 40mm 0mm 40mm 20mm, clip, width=1.\textwidth]{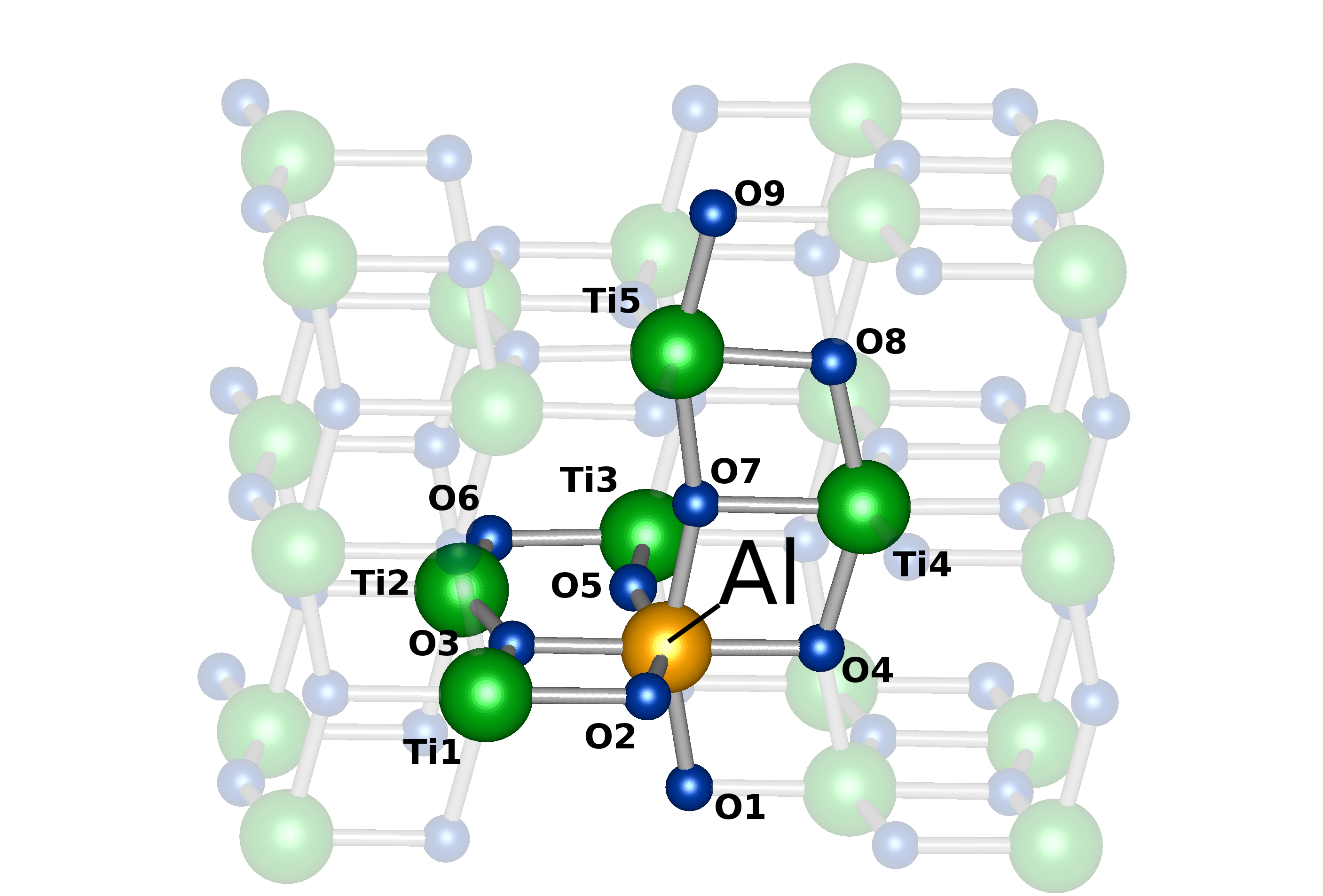}
   \end{minipage}
   &
\begin{minipage}[c]{0.27\textwidth}
    \includegraphics[trim = 40mm 0mm 40mm 20mm, clip, width=1.\textwidth]{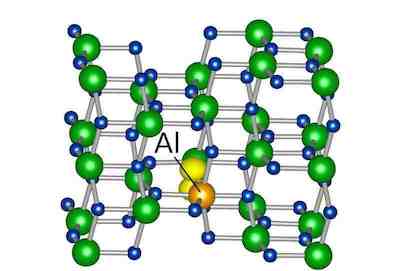}
   \end{minipage}
   &
\begin{minipage}[c]{0.36\textwidth}
\vspace{10mm}
  \includegraphics[angle=-90,origin=c,trim = 0mm 10mm 0mm 10mm, clip, width=1.\textwidth]{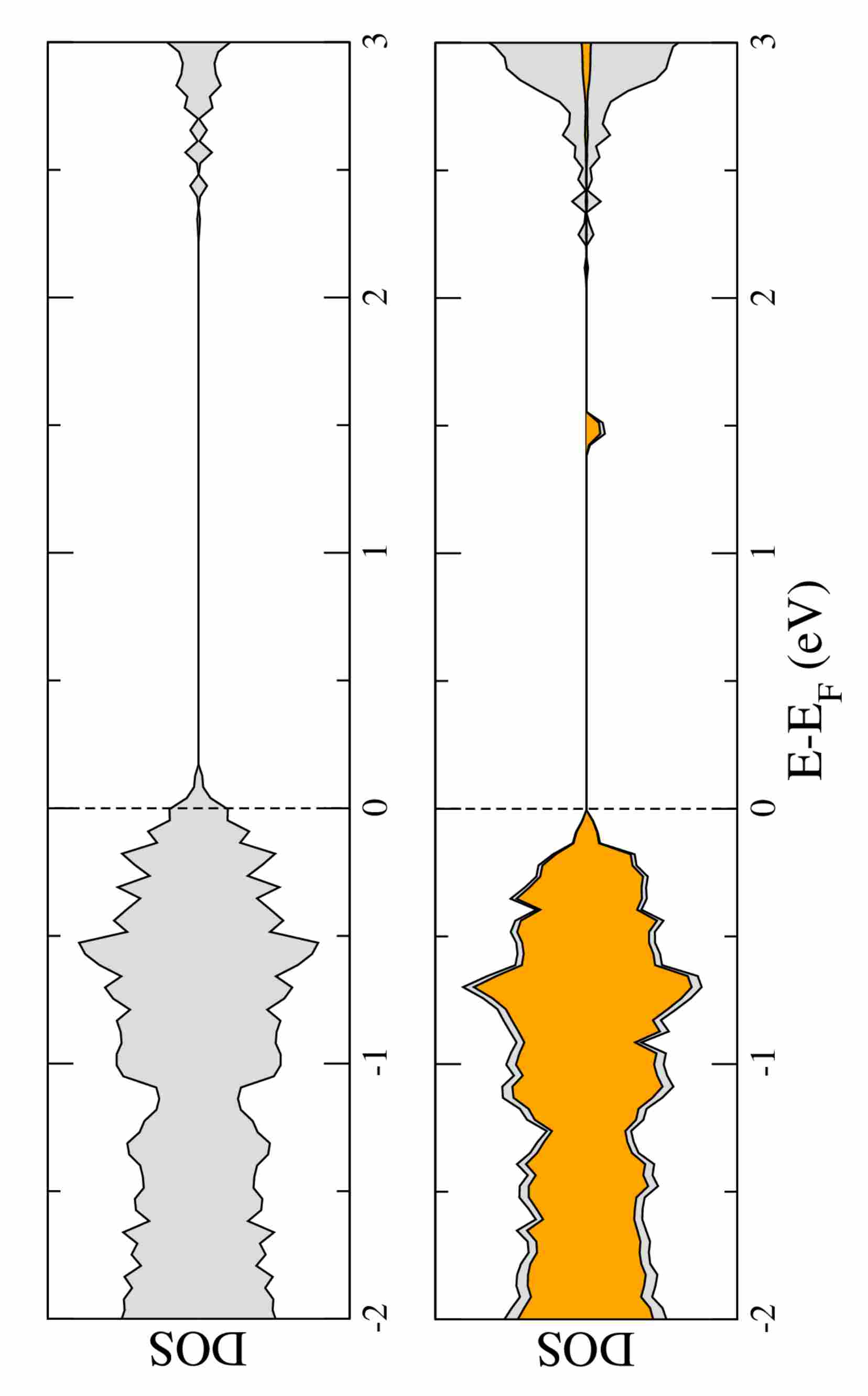}
  \end{minipage}
\\

\textbf{(a)}&\textbf{(b)}&\textbf{(c)}\\
  \end{tabular}
 \end{tabular}
\caption{\textbf{A1 Bulk Defect:} \textbf{Top} Selected bond
lengths for the A1 defect. Largest bond lengths resulting from a
localised polaron are in bold. \textbf{Bottom} \textbf{(a)} Atom labels used in table above. \textbf{(b)} GGA+\emph{U}(6eV) defect state with localised hole. Titaniumm atoms are represented in green, oxygen atoms in blue, aluminium in orange and spin difference isosurface in yellow. 
\textbf{(c)} Projected density of states for A1 defect. Top: GGA,  Bottom: GGA+\emph{U} (\emph{U}=6eV). Total DOS in grey and for GGA+\emph{U} calculations the oxygen 2p states are represented in orange.
}
 \label{A1_SD}
\end{center}
\end{figure}

Defect formation energy and O\superscript{-} hole position
can be seen to have a significant dependence on the value of
the applied \emph{U} correction (no value is 
reported for U=1 eV as convergence was not reached).
Reaction energies vary over a wide range of around 1 eV.
Similar ranges for reactions involving
 TiO\subscript{2} on varying \emph{U} have been reported elsewhere
 \cite{TiO2react}. 
In the case of pure GGA the hole is found to
be delocalised throughout the system, and becomes
increasingly more localised on an oxygen atom neighbouring the
aluminium dopant as the value of \emph{U} is increased.
Hartree-Fock studies of Al doped rutile TiO\subscript{2}
find a well localised polaron associated with the dopant.
Here we find that a \emph{U} value of 6eV provides a well localised hole,
which can be seen in Fig. \ref{A1_SD}, and is close to the value of 7eV
used to describe the polaronic hole in rutile \cite{iwaszuk} and 
oxygen polarons in other materials\cite{dft_pol}.


Selected bond lengths in the
vicinity of the aluminium dopant can be seen in the case of pure
 GGA and GGA+\emph{U} (\emph{U}=6eV). In the case of the pure GGA
calculation, variations in the bond lengths surrounding the dopant
are found to be symmetric. Application of the onsite Coloumb correcton
results in an asymmetric defect. Bond lengths involving an
equatorially bonded oxygen atom adjacent to the dopant are extended, with this 
extension indicative of the associated O\superscript{-} polaron (coordinates
for this structure are provided as supporting information).

Energetically, as this hole becomes more localised with increased
values of \emph{U}, its position is shifted further from the valence
band in the band gap. Partial density of states for the pure
GGA defect and the GGA+\emph{U} (\emph{U}= 6eV) can be seen in
Fig. \ref{A1_SD}(c). For uncorrected GGA the defect is unpolarised with the
oxygen hole located at the top of the valence band, consistent with
results reported by Shirley et al \cite{Shirley-AL-DOP}. For the
GGA+\emph{U} solution the defect is polarised, with the localised
hole in the band gap.

\subsection{A2 \& A3 Defects} 

Calculated bulk energies of reaction for the A2 and A3 defects,
and the comparative result for the A1 defect, are exhibited in 
Table \ref{Bulk-Defect2}. Given the wide variation of calculated reaction energies
for the A1 defect formation with the value of the applied onsite
Coloumb
correction, we have also calculated the reaction energies for the
A2 and A3 defects with an applied correction in order to make
a direct comparison of the stability.

\begin{table}[t]
\begin{center}
\begin{tabular}{c c c c c}
\hline
\hline
Defect Type & GGA &  GGA+\emph{U}  \\
            & (eV) & (\emph{U} = 6eV) \\
\hline
\hline
A1 & 0.293 & -0.472 \\
A2 & -0.388 & -0.710 \\
A3 & -0.317 & -0.629\\
A4 & +2.143 &  N/A  \\
\hline
\hline
\end{tabular}
\caption{Calculated $\Delta\mbox{E}_{0K}$ for bulk aluminium defect types with and without applied \emph{U} correction.
 }
\label{Bulk-Defect2}
\end{center}
\end{table}

GGA predicts defect types comprising two aluminium substitutions with an 
oxygen vacancy, types
A2 \& A3, to be exothermic with the most stable defect type being A2.
Single substitutional defect A1 is found to be endothermic by
GGA. These results are in good agreement with previous work using the
PBE functional, with
ultrasoft pseudopotentials \cite{Shirley-AL-DOP}.
Substitution of an Al\superscript{3+} for one Ti\superscript{4+}
results in one less electron in the system, and a O\superscript{-} state is formed
rather than O\superscript{2-}. Combining two of these substitutions with an oxygen
vacancy results in formal charge being maintained, giving the stability of
defect types A2 \& A3.
While we can put
less faith in the absolute values of the GGA+\emph{U} results, given
the empirical nature of the method and the lack of experimental data,
the same trend is still exhibited with the clustered
 A2 and A3 defect types being more stable than a single Al\superscript{3+} for Ti\superscript{4+} substitution and a similar energy difference between A2 and A3 
defects is found for both GGA and GGA+\emph{U}.

\subsection{A4 Defect}

As a final defect type we have also examined the partial charge compensation of a single dopant combined with an oxygen vacancy. 
 Defect stability for this A4 defect can also be seen in 
table \ref{Bulk-Defect2}, and we see that it is considerably
 less stable than the other three defects. 

Calculations performed using GGA are reported as incorrectly delocalising
oxygen vacancy states throughout the lattice with occupied states
at the bottom of the conduction band\cite{GGA-Ovac}. As such we
also apply the GGA+U method with a value of U=3eV to these states, which has been shown to correctly describe these Ti\superscript{3+} defects 
qualitatively\cite{GGA-Ovac}. In bulk TiO\subscript{2} an oxygen 
vacancy results in the formation of occupied Ti\superscript{3+} 
defect states in the gap.

\begin{figure}
\begin{center}
   \begin{minipage}[t]{0.65\textwidth}
 \includegraphics[angle=-90,trim = 0mm 0mm 0mm 0mm, clip, width=1.\textwidth]{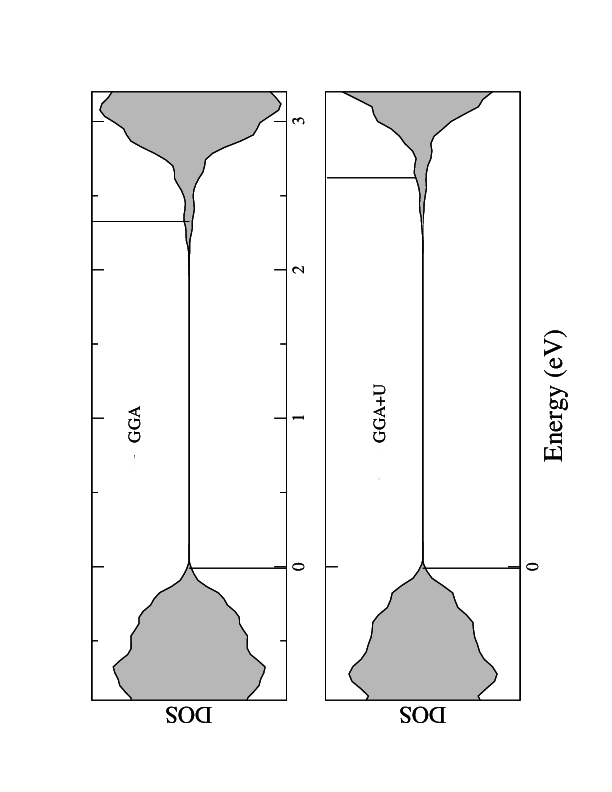}
  \end{minipage}
\caption{Partial density of states for A4 defect. The top of the valence band is at zero, with the highest occupied Kohn-Sham states for each spin 
channel illustrated with a vertical line.}
 \label{A4_PDOS}
\end{center}
\end{figure}

The PDOS for both the pure
GGA and the GGA+U solutions can be seen in Fig. \ref{A4_PDOS}
Both methods produce an occupied vacancy defect state at
the bottom of the conduction band, which is
delocalised throughout the lattice, while the oxygen
hole associated with the Al dopant has been removed.
This partial charge compensation can be seen as the 
most simple arrangement of an Al and vacancy defect leading 
to the clean up of Ti\superscript{3+} states.

\section{(101) Surface}

\begin{figure}
\begin{center}
   \begin{minipage}[t]{0.45\textwidth}
 \includegraphics[trim = 0mm 0mm 0mm 0mm, clip, width=1.\textwidth]{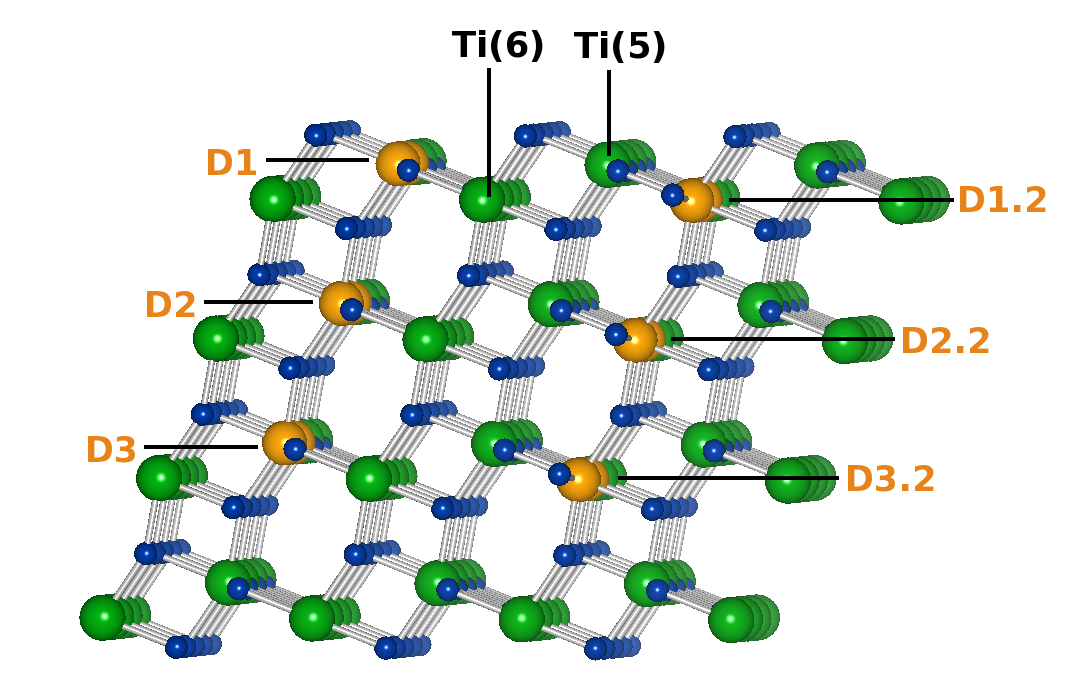}
  \end{minipage}
\caption{Notation for surface defect positions relative to the 101 surface.
 Titanium atoms are represented in green, oxygen atoms in blue and 
aluminium in orange. Ti(5) labels five-fold co-ordinated titanium atoms,
 Ti(6) six-fold coordinated.}
 \label{Defects2}
\end{center}
\end{figure}

In the case of the (101) surface numerous inequivalent positions are available 
for defects. The clean stoichiometric surface contains both five-fold and six-fold
co-ordinated titanium atoms, each of which may be substituted for an Al atom. 
Energies of reaction for A2 and A3 defects in which substitutions occurred in 
different (101) layers were consistently found to be less stable than those 
containing two Al atoms in the same layer and therefore we only report the latter 
results. Differing defect positions for A2 defects can be seen in
 Fig. \ref{Defects2}. For the A1 defect position the same notation applies, for example 
D1 corresponds to a single substitution of a five-fold co-ordinated Ti atom and 
D1.2 corresponds to substitution of a six-fold co-ordinated Ti in the upper layer.
 A3 defects 
necessarily contain Al substitutions at slightly differing positions along the [101] 
direction, which in the case of the uppermost layer means substitution of
 one five-fold co-ordinated and one six-fold co-ordinated Ti. The differing 
notation therefore refers to the position of the vacancy, with D1 refering to 
the defect in the uppermost layer with the vacancy slightly further along the [101]
direction than D1.2.

Calculated reaction energies for defects at each of these positions are shown in 
table \ref{Surf-Defect}. It is known that oxygen vacancies on anatase (101) surfaces
reside preferentially at sub-surface sites \cite{Selloni-VAC1,Selloni-VAC2}.
Sub-surface sites are here also found to be preferential for the 
aluminium defects, with
the D2 position being the most stable for each type (the same was found in tests 
with 5 layer slabs). 
Large differences in stability can be seen between surface and subsurface sites, 
with energetic bias towards the sub-surface positions as much as 0.85 eV. 
Reaction energies tend to converge as the defect moves further into the
bulk, and the highest energies are present when aluminium atoms reside next
to two-fold cordinated oxygen atom, O(2), in the surface i.e. position D1.2.
Despite residing next to an O(2) atom the A2 defect at position D1 is relatively 
more stable than its D1.2 counterpart as the Al atoms bonding to the 
O(2) atoms are themselves under co-ordinated as a result of being at the surface.


\begin{table}[t]
\begin{center}
\begin{tabular}{c c c c c c}
\hline
\hline
Defect Type & \multicolumn{2}{c}{A1} & A2 & A3 & A4\\
            & GGA  & GGA+U & GGA  & GGA & GGA\\
            & (eV) & (eV) & (eV) & (eV) & (eV) \\
\hline
Position   & & & & &\\
\hline
D1     &  0.197 & -0.056 & -0.128  & 0.274 & 2.097 \\
D1.2   &  0.454 &  0.333 & 0.625  & 0.331 & 2.514 \\
D2     & \textbf{-0.122} & \textbf{-0.225} & \textbf{-0.568} & \textbf{-0.337} & \textbf{1.663} \\
D2.2   &  0.043 &  0.094 & -0.175  &-0.299 & 1.931 \\
D3     & -0.050 &  0.164 & -0.400  &-0.299 & 1.800 \\
D3.2   & -0.020 &  0.328 & -0.307  &-0.301 & 1.854  \\
\hline
\hline
\end{tabular}
\caption{Calculated $\Delta\mbox{E}_{0K}$ for differing positions in a 4 layer (101) slab. Most stable defect positions for each type are highlighted in bold. A1 GGA+\emph{U} reaction energies are given relative to the bulk defect formation energy. }
\label{Surf-Defect}
\end{center}
\end{table}

\begin{figure}[t]
\begin{center}

\begin{tabular}{c c c}
  \begin{minipage}[c]{0.3\textwidth}
  \includegraphics[angle=0,origin=c,trim = 50mm 10mm 25mm 0mm, clip, width=1.\textwidth]{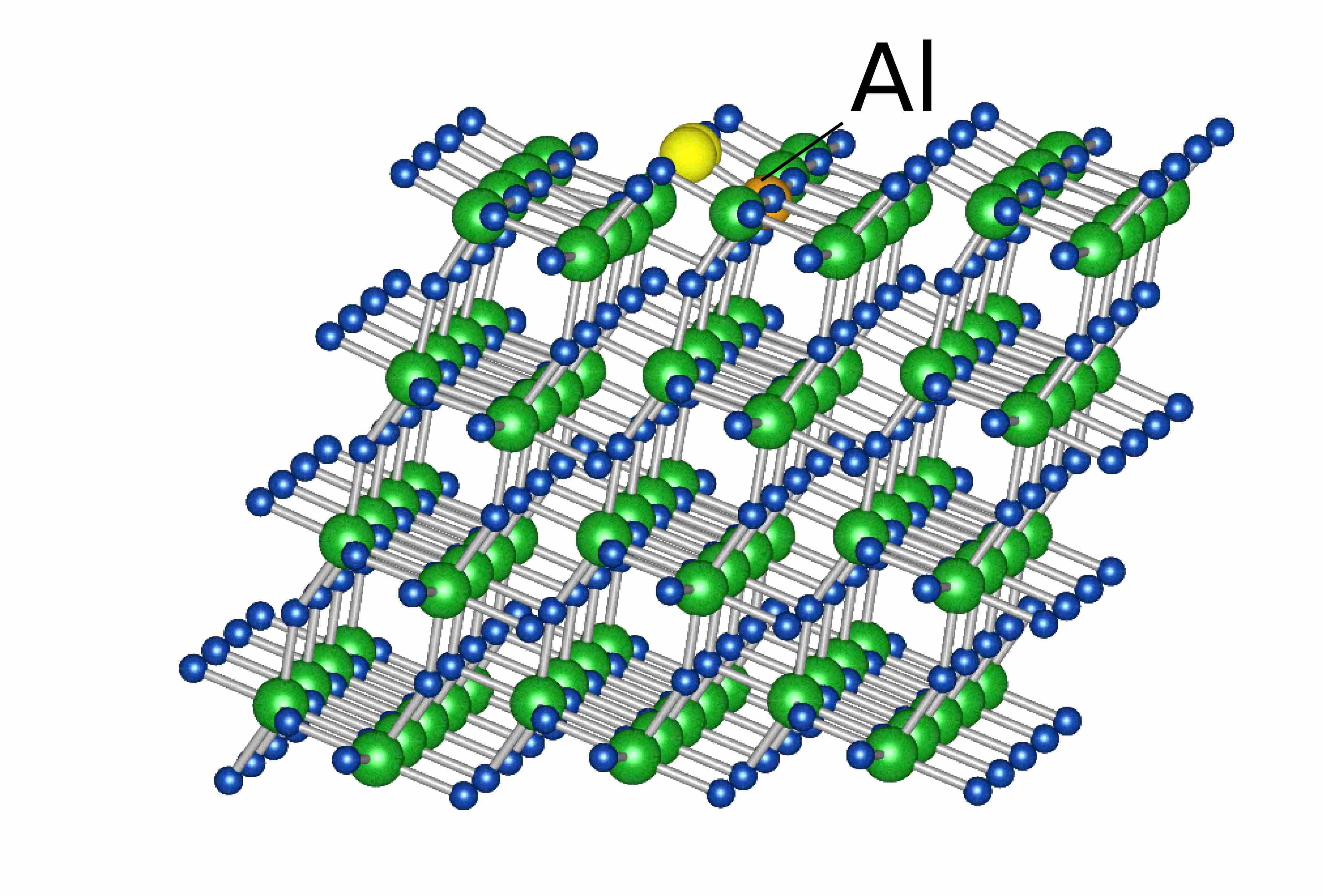}
  \end{minipage}&
  \begin{minipage}[c]{0.3\textwidth}
  \includegraphics[angle=0,origin=c,trim = 50mm 10mm 25mm 0mm, clip, width=1.\textwidth]{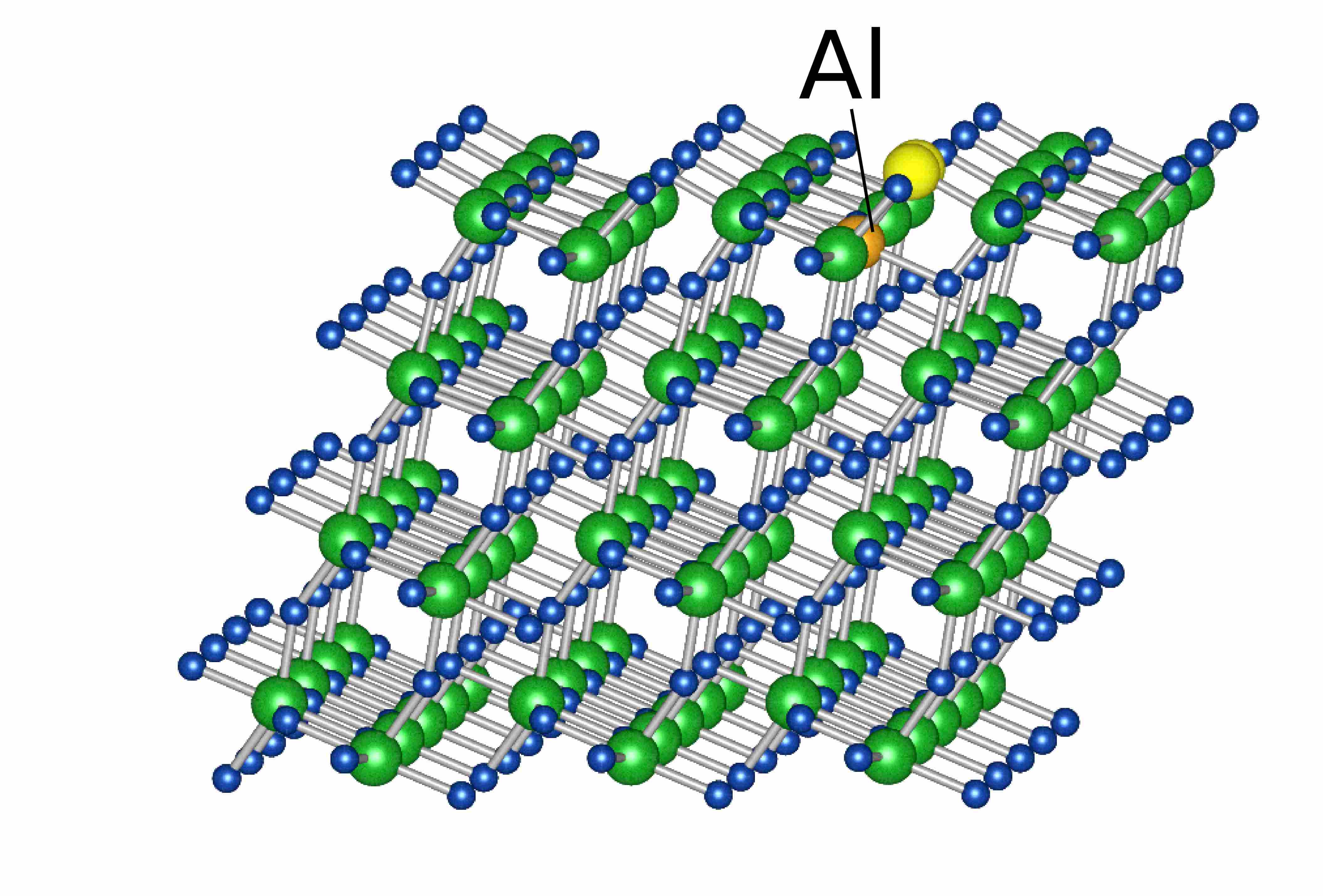}
  \end{minipage}&
  \begin{minipage}[c]{0.3\textwidth}
  \includegraphics[angle=0,origin=c,trim = 50mm 10mm 25mm 0mm, clip, width=1.\textwidth]{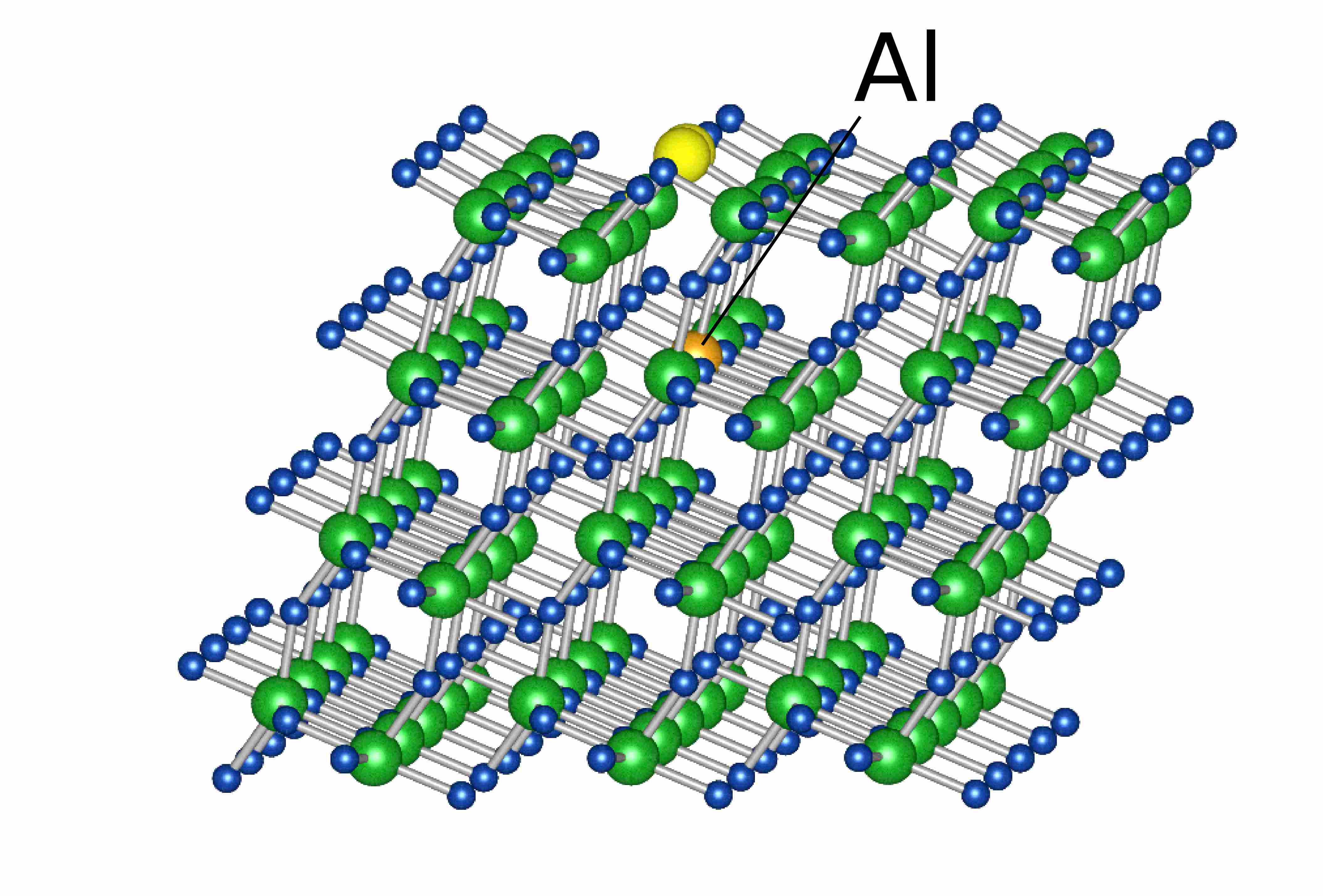}
  \end{minipage}\\
\textbf{D1:} $E_{1}= 0.154\textrm{ eV}$ & \textbf{D2.2:} $E_{1}=0.321\textrm{ eV}$ & \textbf{D2} $E_{1}=0.709\textrm{ eV}$\\
  \begin{minipage}[c]{0.3\textwidth}
  \includegraphics[angle=0,origin=c,trim = 50mm 10mm 25mm 0mm, clip, width=1.\textwidth]{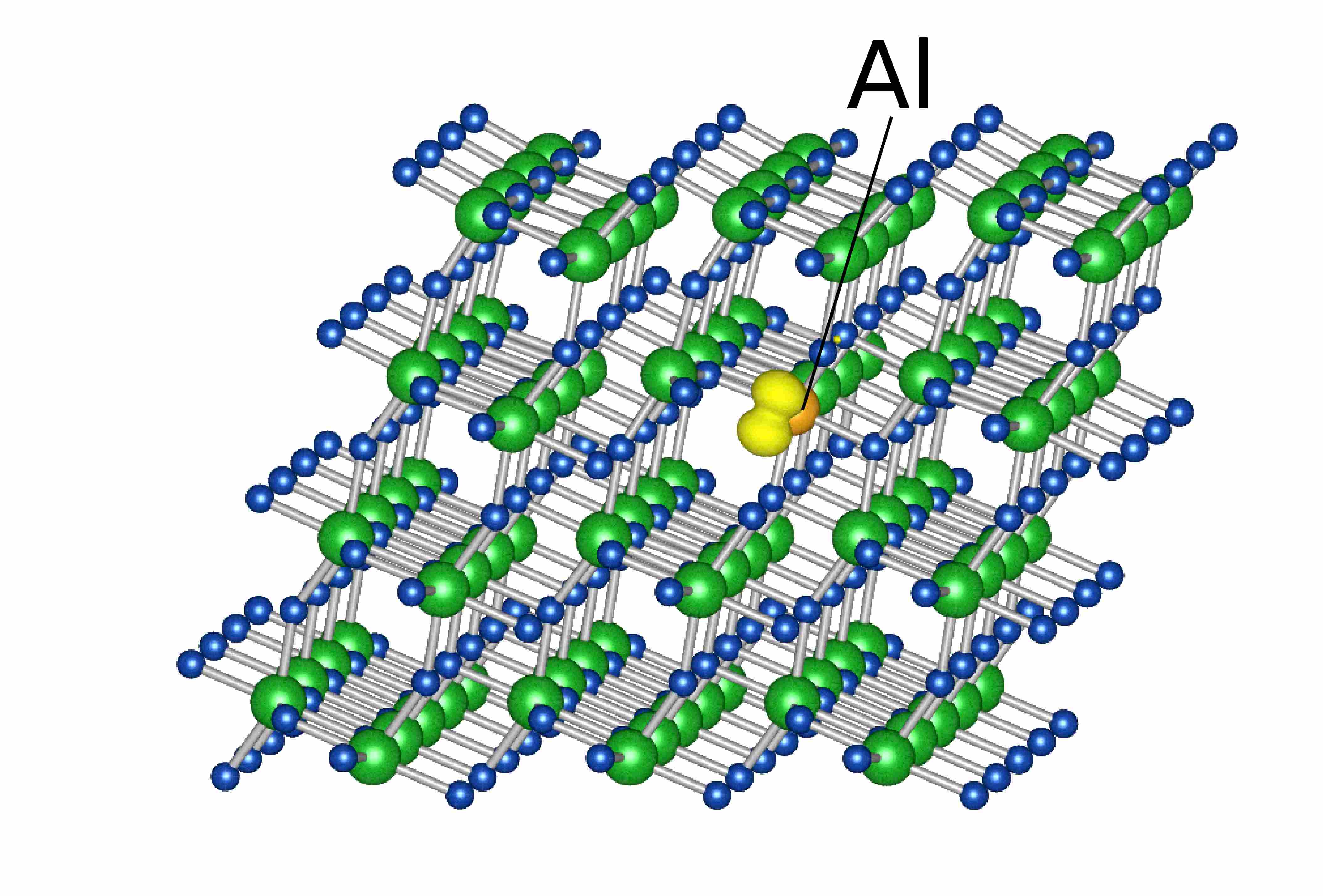}
  \end{minipage}&
  \begin{minipage}[c]{0.3\textwidth}
  \includegraphics[angle=0,origin=c,trim = 50mm 10mm 25mm 0mm, clip, width=1.\textwidth]{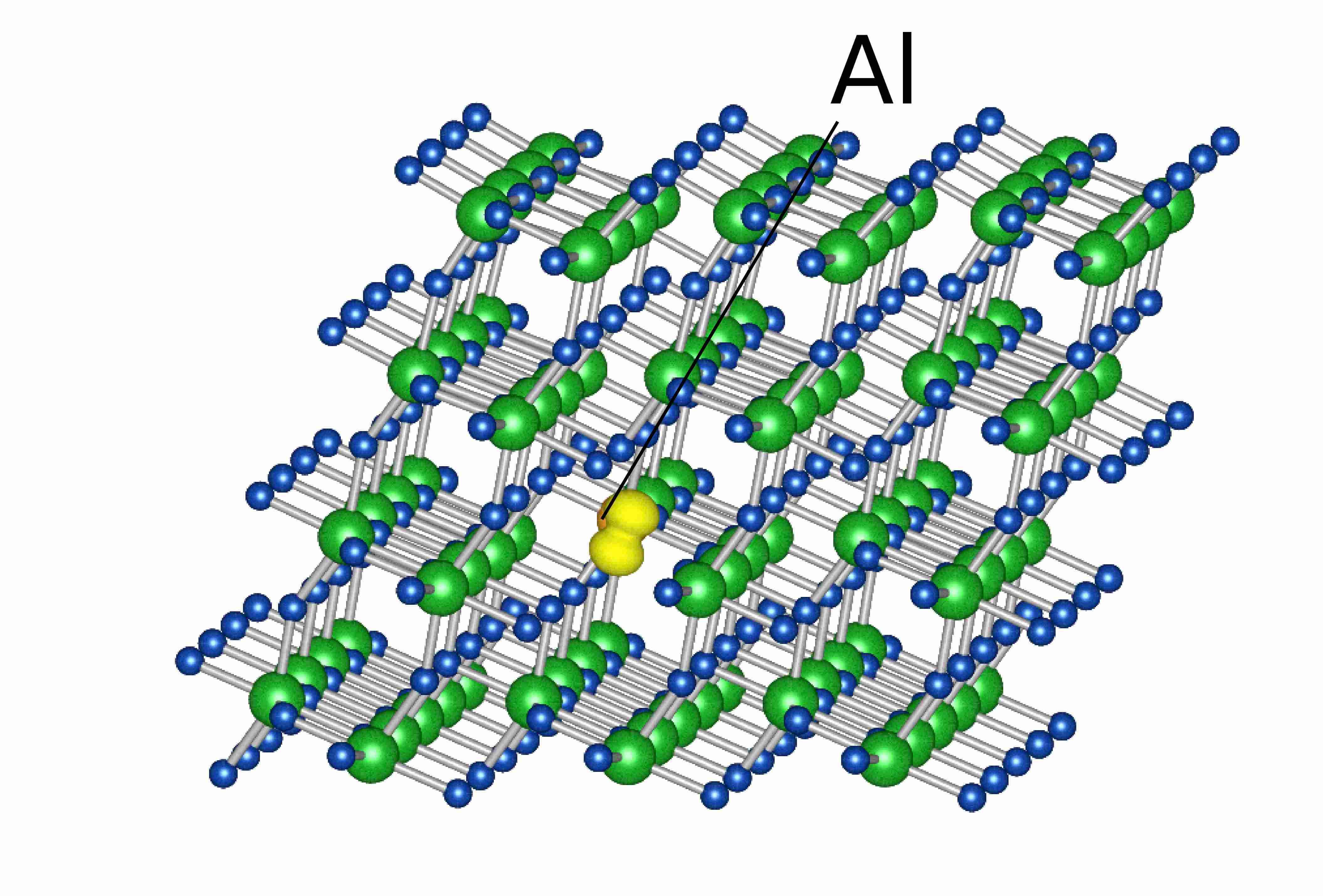}
  \end{minipage}&
  \begin{minipage}[c]{0.3\textwidth}
  \includegraphics[angle=0,origin=c,trim = 50mm 10mm 25mm 0mm, clip, width=1.\textwidth]{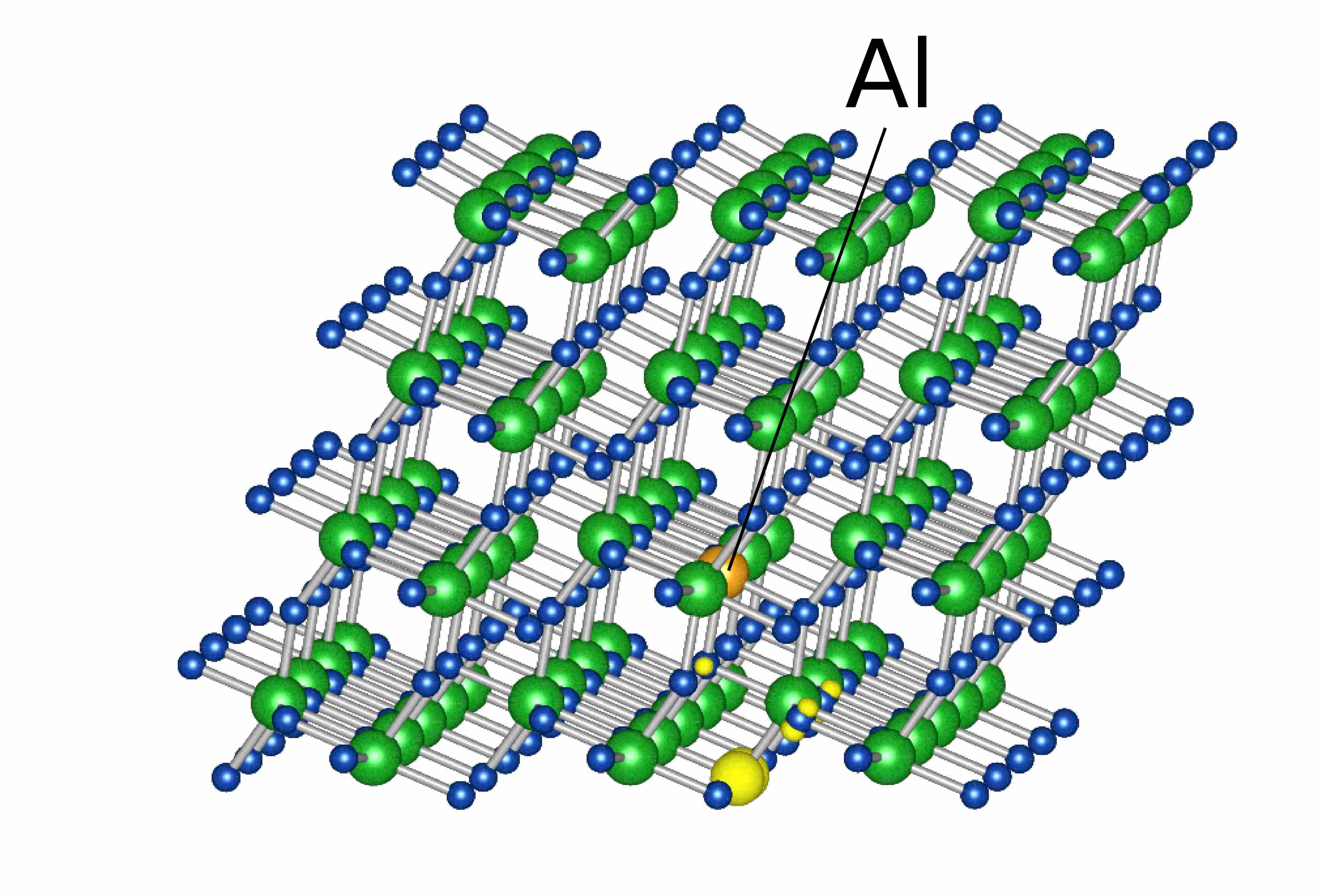}
  \end{minipage}\\
\textbf{D2.2:} $E_{1}=0.586\textrm{ eV}$ & \textbf{D3:} $E_{1}=0.661\textrm{ eV}$ & \textbf{D3.2:} $E_{1}=1.78\textrm{ eV}$\\
\end{tabular}

\caption{\textbf{A1 Surface Defect:} GGA+\emph{U} spin difference
isosurface 
for the A1 single substitutional defect at varying positions in the (101)
surface. Associated hole position relative to the conduction
band is also given.}
 \label{A1_SURF}
\end{center}
\end{figure}

Given the empirical nature of the GGA+\emph{U} reaction energies,
the A1 GGA+\emph{U} results are given relative to the bulk case
in order to focus on the general trend rather than absolute values.
Similar energetic bias towards the subsurface D2 is exhibited,
with the D1.2 position the least stable.
GGA+\emph{U} results for the A1 defect suggest that if the defect
resides close enough to an under co-ordinated surface O(2) atom
the O\superscript{-} hole will localise on it. Defect positions
D1, D1.2 and D2 all result with a surface localised hole state
, as can be seen in Fig. \ref{A1_SURF}.
Indeed in the case of the D3.2 defect the hole localises on
an O(2) atom in the bottom layer of the surface which, given that the
bottom layer of the slab is fixed to the bulk position, explains
the increase in the D3.2 defect formation energy relative to the bulk.

\begin{figure}
\begin{center}
   \begin{minipage}[t]{0.85\textwidth}
 \includegraphics[angle=-90,origin=c,trim = 60mm 30mm 30mm 60mm, clip, width=1.\textwidth]{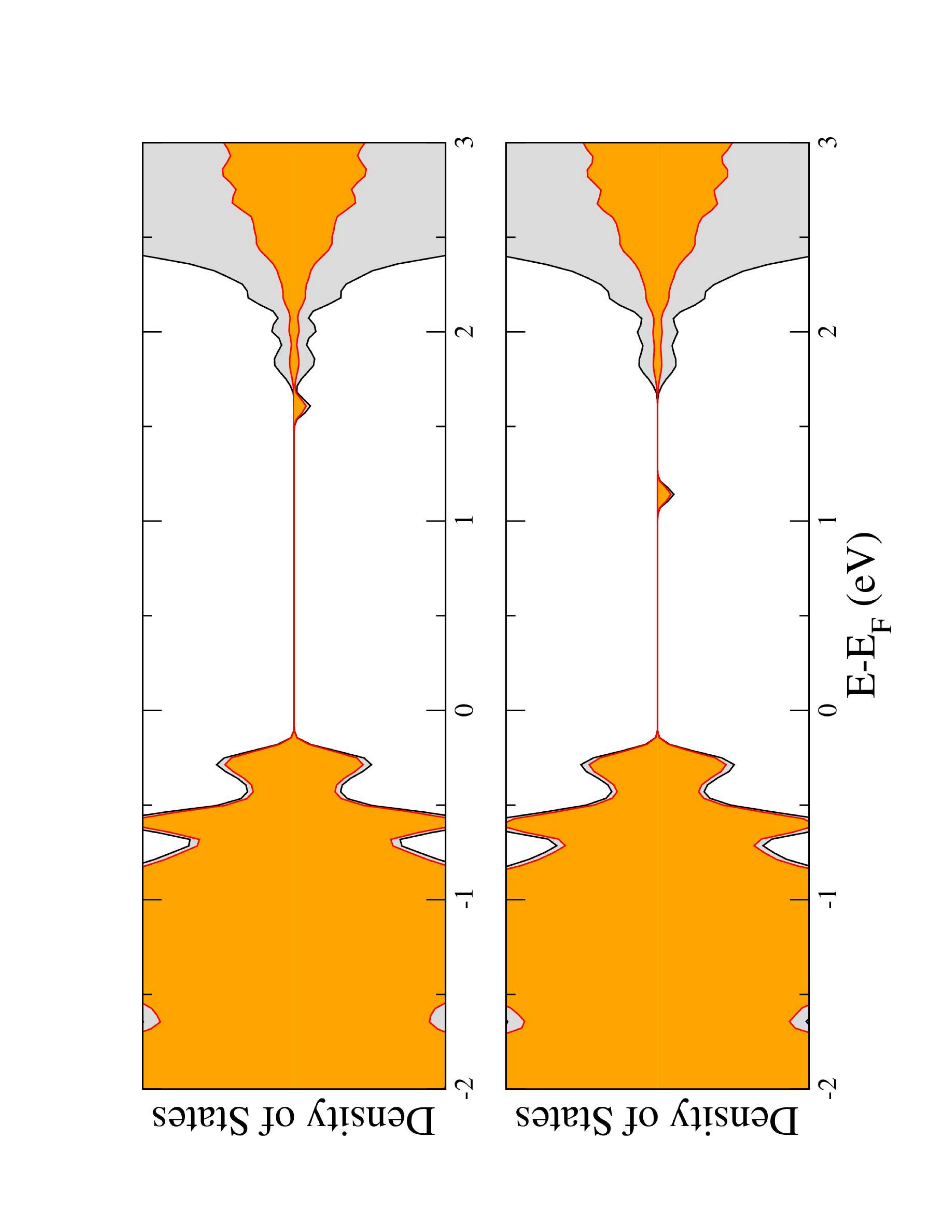}
  \end{minipage}
\caption{Partial density of states for single Al dopants in the (101) surface. Top: A1 defect in D1 surface position. Bottom: A1 defect in D2.2 position. Projection on the oxygen atoms is in orange, total DOS in grey.}
 \label{A1_PDOS}
\end{center}
\end{figure}

Energetically GGA predicts no variation in position relative to the
valence band of the hole state, while for the GGA+\emph{U}
calculations a large energetic variation is observed with
differing location in the surface Fig. \ref{A1_SURF}. 
At the surface the hole state is found energetically further 
from the valence band, converging towards the
bulk value as it moves away from the surface into the slab. This effect
may be seen for
two examples in Fig. \ref{A1_PDOS} (D3.2 being the exception
again, due to the localisation of the hole on a fixed atom).

As in the bulk case, the most stable defect is found to be that of type A2 and its formation is energetically favourable when at the D2 sub-surface position. Similarly, as in the bulk case, defect A4 is significantly less stable than the other three defect types. 
An important difference to note is that GGA predicts that
when A1 defects reside at subsurface sites the defect formation
becomes exothermic and energetically favourable, unlike the bulk
which is endothermic. It is worth noting that the differences 
between all three defect stabilities are not so large, with $\sim$ 0.2 eV 
differences between the calculated stabilities for the A1 and A3 defects, 
and the A3 and A2 defects respectively.

 \subsection{Oxygen Vacancy}

\begin{figure}
\begin{center}
\begin{tabular}{c c}
   \begin{minipage}{0.35\textwidth}
 \includegraphics[trim = 0mm 0mm 0mm 0mm, clip, width=1.\textwidth]{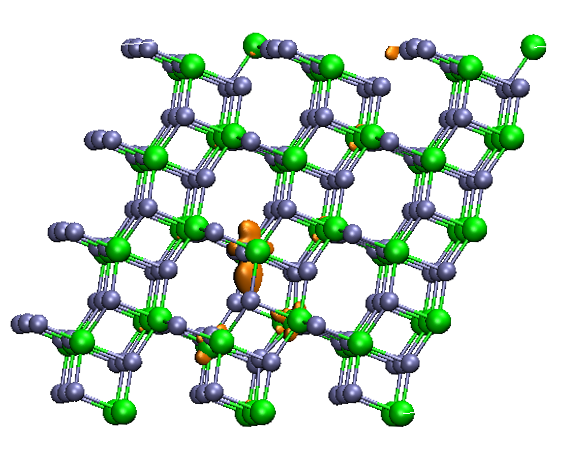}
    \end{minipage}
&
\begin{minipage}{0.65\textwidth}
 \includegraphics[angle=-90,trim = 60mm 20mm 20mm 40mm, clip, width=1.\textwidth]{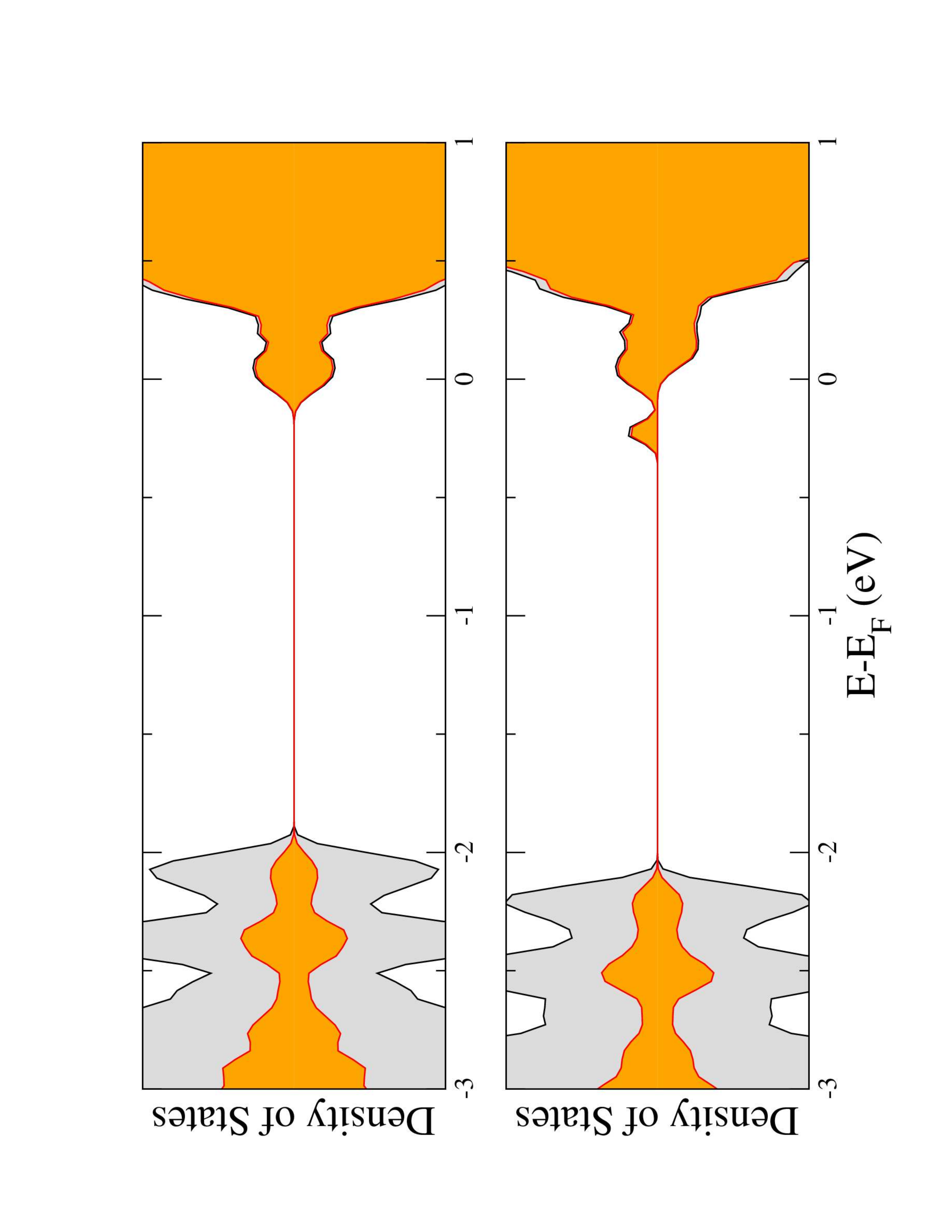}
    \end{minipage}\\
\textbf{(a)}&\textbf{(b)}
\end{tabular}
\caption{\textbf{(a)} Spin isosurface for oxygen vacancy state in the (101) surface and \textbf{(b)}partial density of states for GGA(top) and GGA+U(U=3eV)(bottom)  }

 \label{Ovac}
\end{center}
\end{figure}

We have seen that introduction of an Al dopant produces hole states
in the band gap. Intrinsic oxygen vacancies have the opposite
effect, introducing occupied states
within
the band gap. In the case of GGA these states are delocalised
throughout the lattice, and unpolarised with the introduced
states both at the bottom of the conduction band.
Application of DFT+U produces
a localised vacancy state on a neighbouring Ti atom, giving us a
Ti\superscript{3+} ion, and a second delocalised throughout
the lattice (spin difference isodensity plot may be seen in
Fig. \ref{Ovac}). We can see from the
partial density of states that both states lie relatively
close to the conduction band edge, with the localised
state further into the band gap.
This is similar to the situation
reported for oxygen vacancies in bulk anatase treated
with GGA+U \cite{GGA-Ovac}, however in the bulk two different 
solutions are found for the U=3eV case, one with a localised vacancy state
and a delocalised vacancy state and a second in which both states
are localised. Here we find only one stable structure, which we assume is
due to the surface perturbation of the structure relative to that of the bulk.

%
%
%

\section{Vacancy Diffusion}

Thus far we have examined several defects in which extrinsic dopants
cluster with intrinsic oxygen vacancies without discussing how this
clustering will occur. Previous work has suggested that the movement
of aluminium interstitials throughout the anatase lattice is not
energetically feasible\cite{Shirley-AL-DOP}. 
Here we propose the diffusion of oxygen vacancies throughout
the lattice as a mechanism by which Al dopants and oxygen vacancies
can combine.

\begin{figure}[t]
\begin{tabular}{c c}
\begin{minipage}[t]{0.45\textwidth}
\hspace{-0.1\textwidth}
\begin{tabular}{c l c l c}
  \begin{minipage}[t]{0.31\textwidth}
  \includegraphics[trim = 300mm 10mm 350mm 50mm, clip, width=1.2\textwidth]{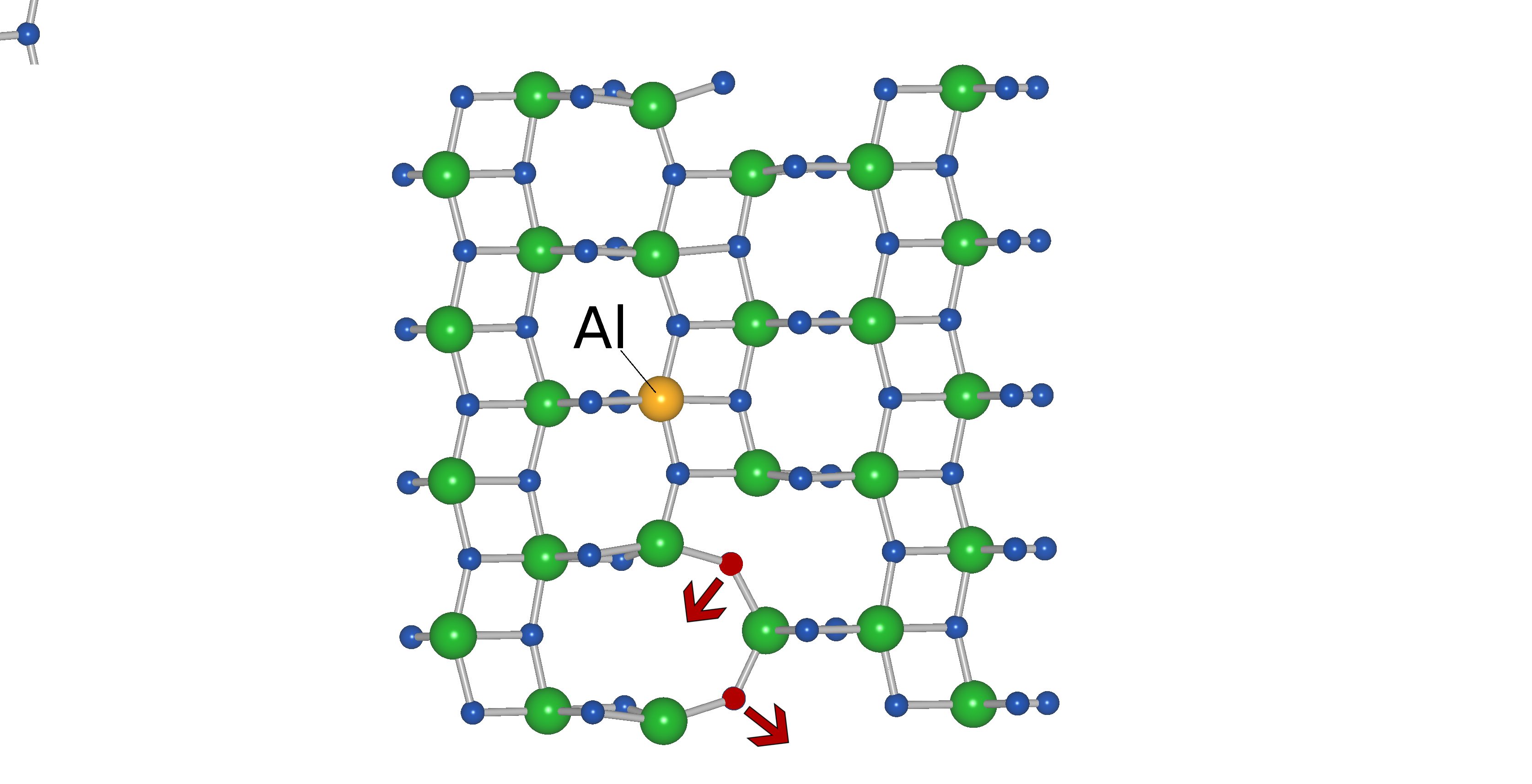}
  \end{minipage}
  &
 \begin{minipage}[t]{0.025\textwidth}
 \vspace{-0.6in}
    \includegraphics[angle=-90,origin=c,trim = 0mm 0mm 0mm 0mm, clip,width=1.2\textwidth]{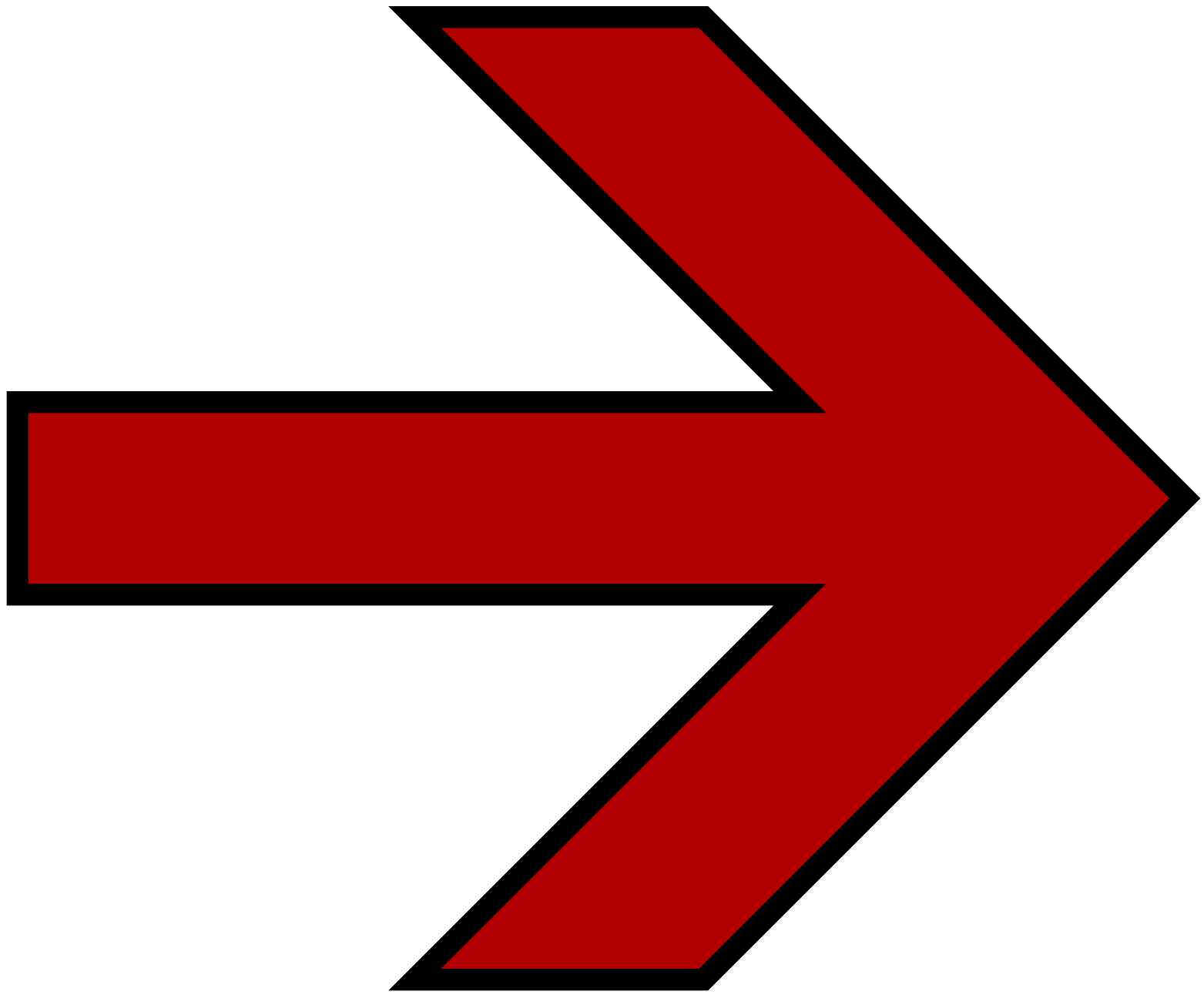}
  \end{minipage}&
  \begin{minipage}[t]{0.31\textwidth}
 \includegraphics[trim = 300mm 10mm 350mm 50mm, clip, width=1.2\textwidth]{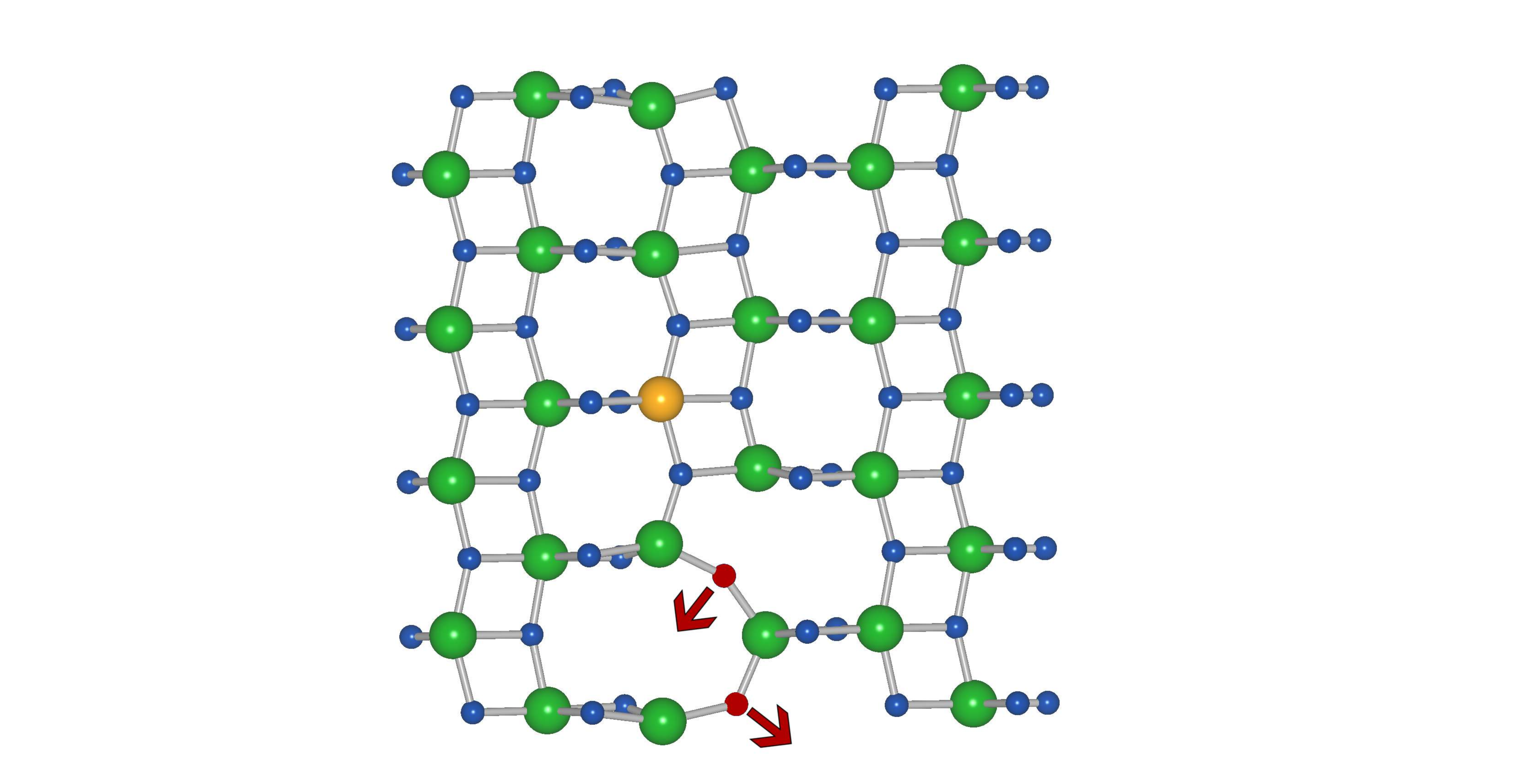}
  \end{minipage}
  &
 \begin{minipage}[t]{0.025\textwidth}
 \vspace{-0.6in}
    \includegraphics[angle=-90,origin=c,trim = 0mm 0mm 0mm 0mm, clip,width=1.2\textwidth]{IMAGES/arrow.pdf}
  \end{minipage}&
  \begin{minipage}[t]{0.31\textwidth}
 \includegraphics[trim = 300mm 10mm 350mm 50mm, clip, width=1.2\textwidth]{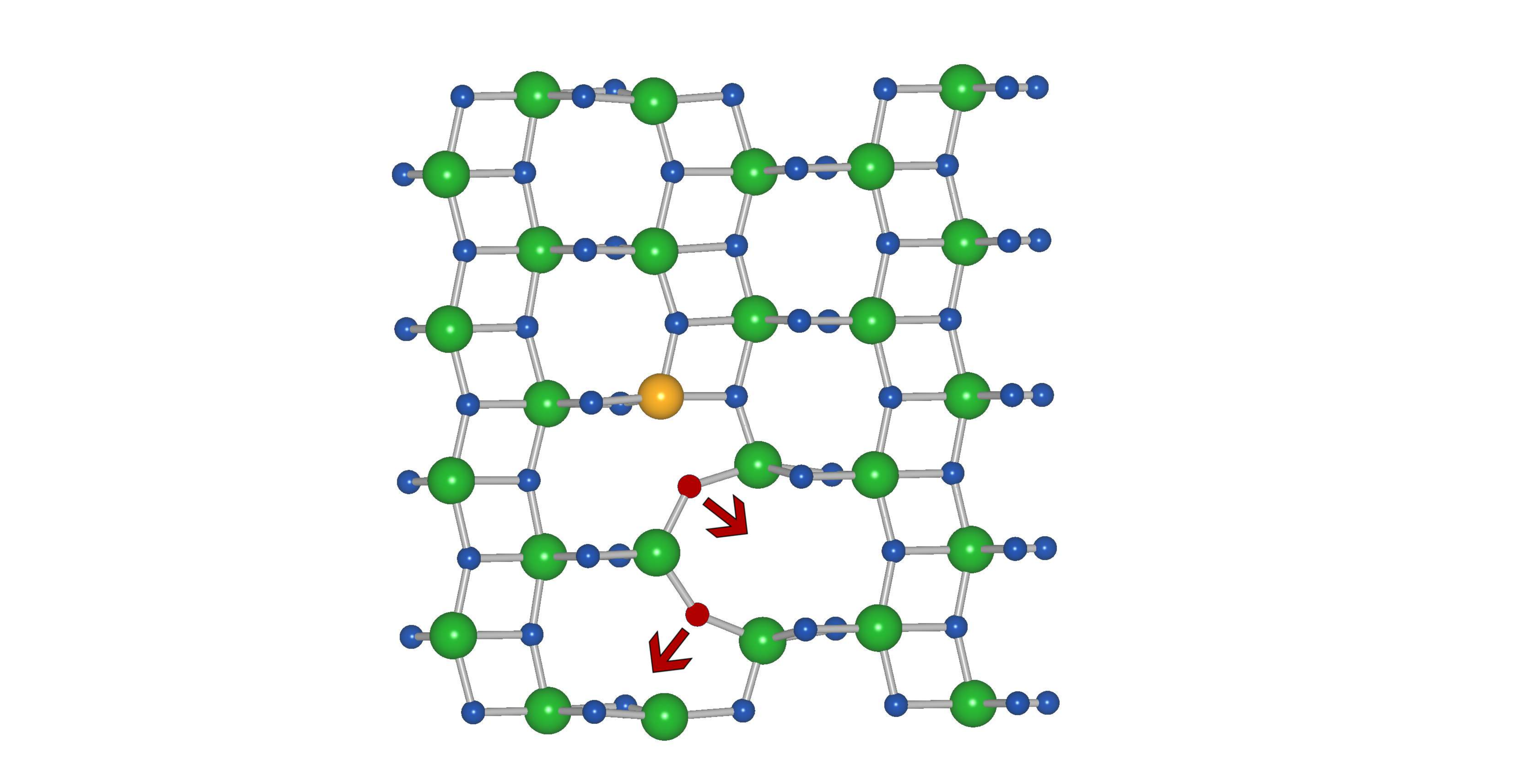}
  \end{minipage}\\
 0&&2&&5\\
&&&&\begin{minipage}[t]{0.025\textwidth}
    \rotatebox[origin=c]{0}{\includegraphics[angle=180,origin=c,trim = 0mm 0mm 0mm 0mm, clip,totalheight=1.2\textwidth]{IMAGES/arrow.pdf}}
  \end{minipage}\\

  \begin{minipage}[t]{0.31\textwidth}
 \includegraphics[trim = 300mm 10mm 350mm 50mm, clip, width=1.2\textwidth]{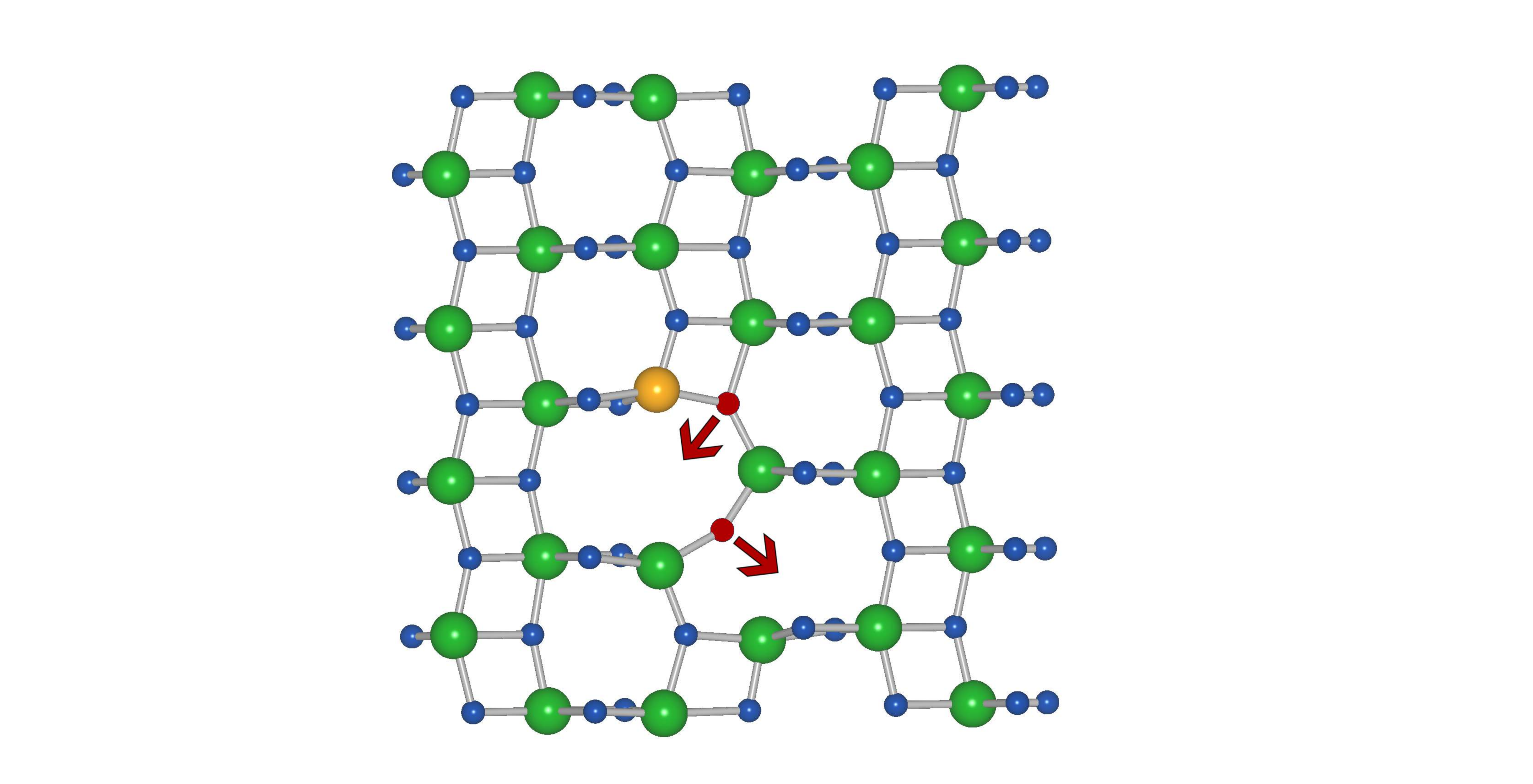}
  \end{minipage}
  &
 \begin{minipage}[t]{0.025\textwidth}
 \vspace{-0.6in}
    \includegraphics[angle=90,origin=c,trim = 0mm 0mm 0mm 0mm, clip,width=1.2\textwidth]{IMAGES/arrow.pdf}
  \end{minipage}&
  \begin{minipage}[t]{0.31\textwidth}
 \includegraphics[trim = 300mm 10mm 350mm 50mm, clip, width=1.2\textwidth]{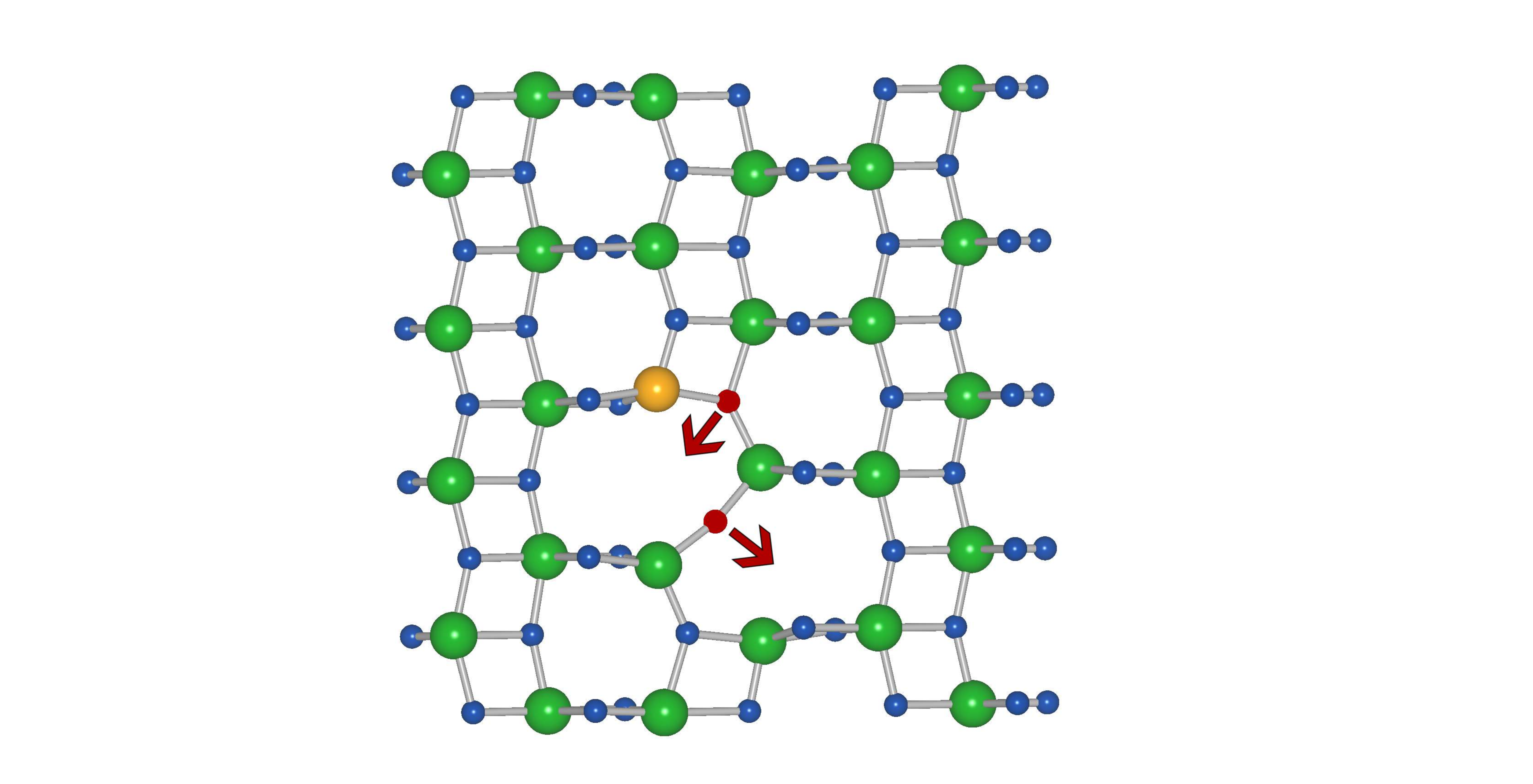}
  \end{minipage}
  &
 \begin{minipage}[t]{0.025\textwidth}
 \vspace{-0.6in}
    \includegraphics[angle=90,origin=c,trim = 0mm 0mm 0mm 0mm, clip,width=1.2\textwidth]{IMAGES/arrow.pdf}
  \end{minipage}&
  \begin{minipage}[t]{0.31\textwidth}
 \includegraphics[trim = 300mm 10mm 350mm 50mm, clip, width=1.2\textwidth]{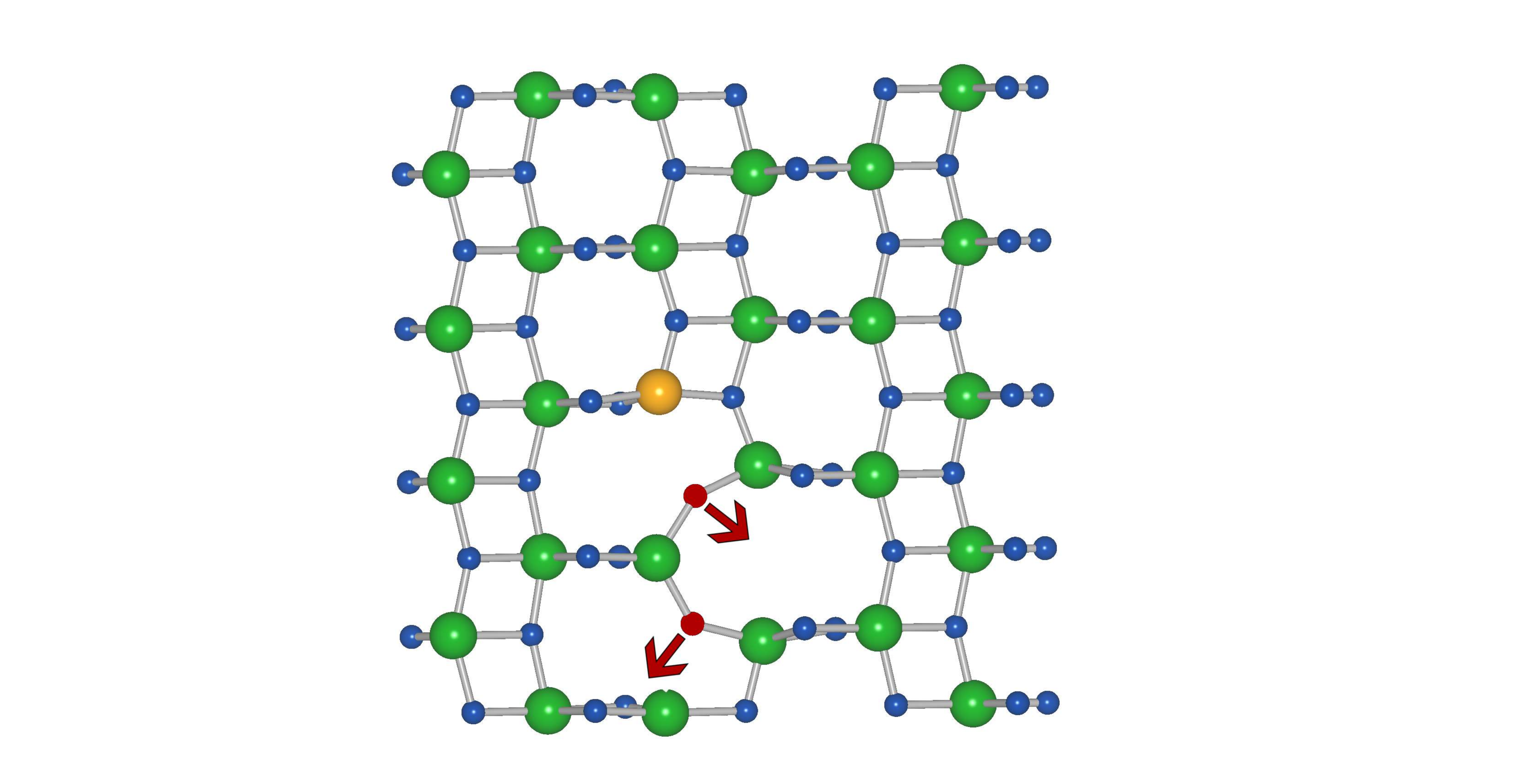}
  \end{minipage}\\
9&&8&&6\\

\end{tabular}
\end{minipage}
&
\hspace{0.05\textwidth}
\begin{minipage}[c]{0.5\textwidth}
\vspace{-0.1\textwidth} \includegraphics[width=1.\textwidth]{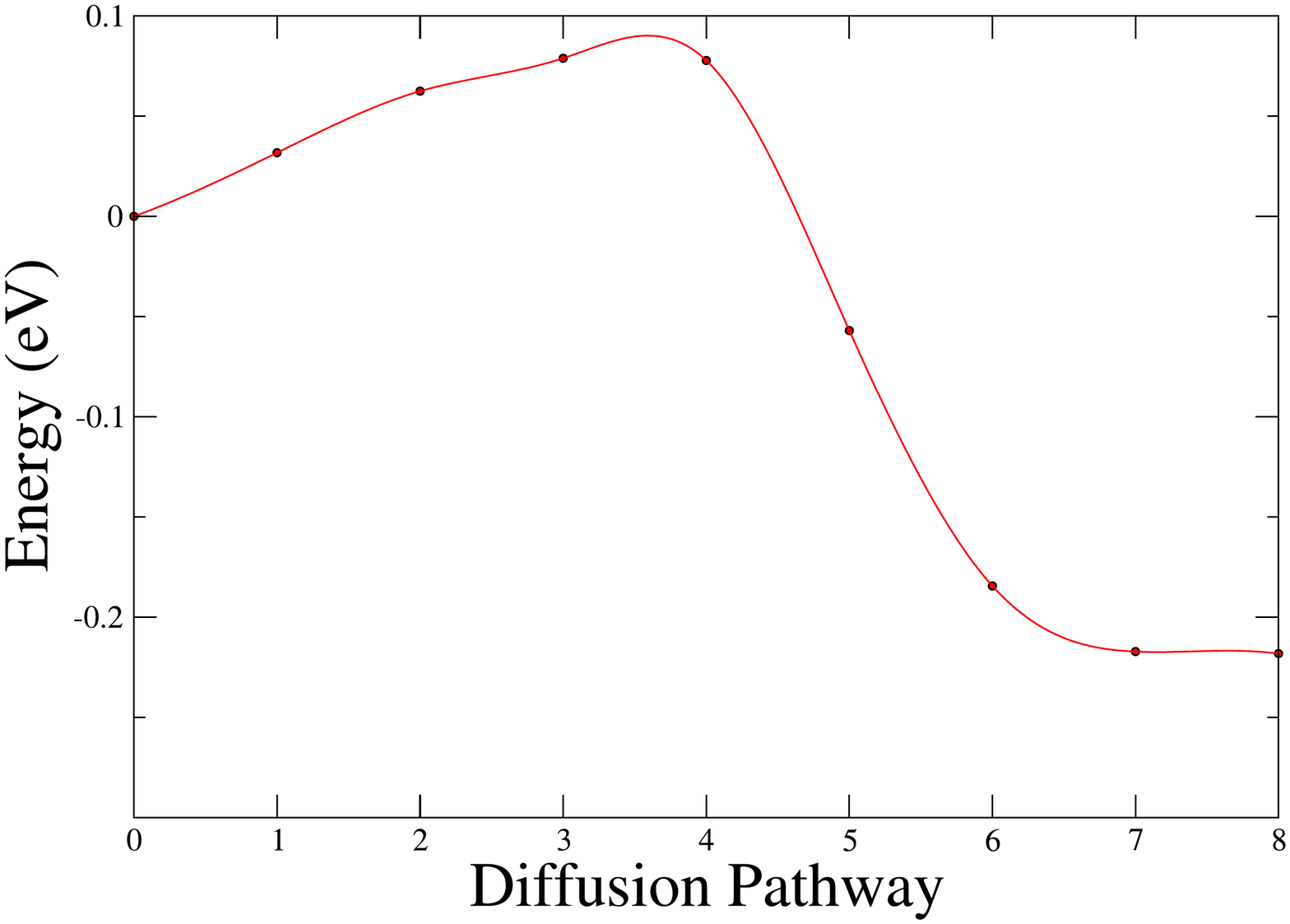}
\end{minipage}\\
\textbf{(a)}&\textbf{(b)}\\
\end{tabular}
\caption{\textbf{(a)} Diffusion pathway for an oxygen vacancy towards Aluminium dopant resulting in a defect of A4 type. For clarity only the central layer is illustrated. \textbf{(b)} Potential energy pathway along the shown diffusion pathway, with a spline fitted to the data to serve as a guide to the eye.}
 \label{A4-diff}
\end{figure}

In order to investigate this possibility we have performed climbing image
nudged elastic band (NEB)
calculations\cite{Henkelman-NEB}, using the GGA functional, to examine
the energy profile along oxygen vacancy diffusion
paths which would result in the formation of A2, A3 and A4 defects.
For these calculations the computational expense has been eased by reducing the number
of anatase layers in the slab to three (sub-surface defects in position D2, i.e. those in
the second layer from the surface, remain the most energetically favoured 
in the three layer slab). Given the preference of oxygen vacancies to reside subsurface
in anatase TiO\subscript{2} and their
reported mobility\cite{cheng2}, we examine the diffusion of these 
subsurface oxygen vacancies towards the aluminium dopants, with diffusion
proceeding parallel to the (101) surface. This is taken as a viable 
mechanism for the combination of vacancies with dopants, given their
preference to reside sub-surface, although it is also 
entirely possible that vacancies could diffuse directly from the
surface towards Al dopants. 

Firstly a single oxygen vacancy diffusing towards a single A1 defect to form the A4 defect is examined, with the vacancy pathway and energy profile 
illustrated in Fig. \ref{A4-diff}.
 A clear
enegetic bias ($\sim$ 0.2 eV) towards the formation of the A4 defect is 
exhibited, with an
extremely small barrier of around 0.08 eV, which will be easily overcome
at room temperature.

\begin{figure}[t]
\begin{tabular}{c c}
\begin{minipage}[t]{0.45\textwidth}
\hspace{-0.1\textwidth}
\begin{tabular}{c l c l c}
  \begin{minipage}[t]{0.31\textwidth}
  \includegraphics[trim = 20mm 0mm 20mm 0mm, clip, width=1.2\textwidth]{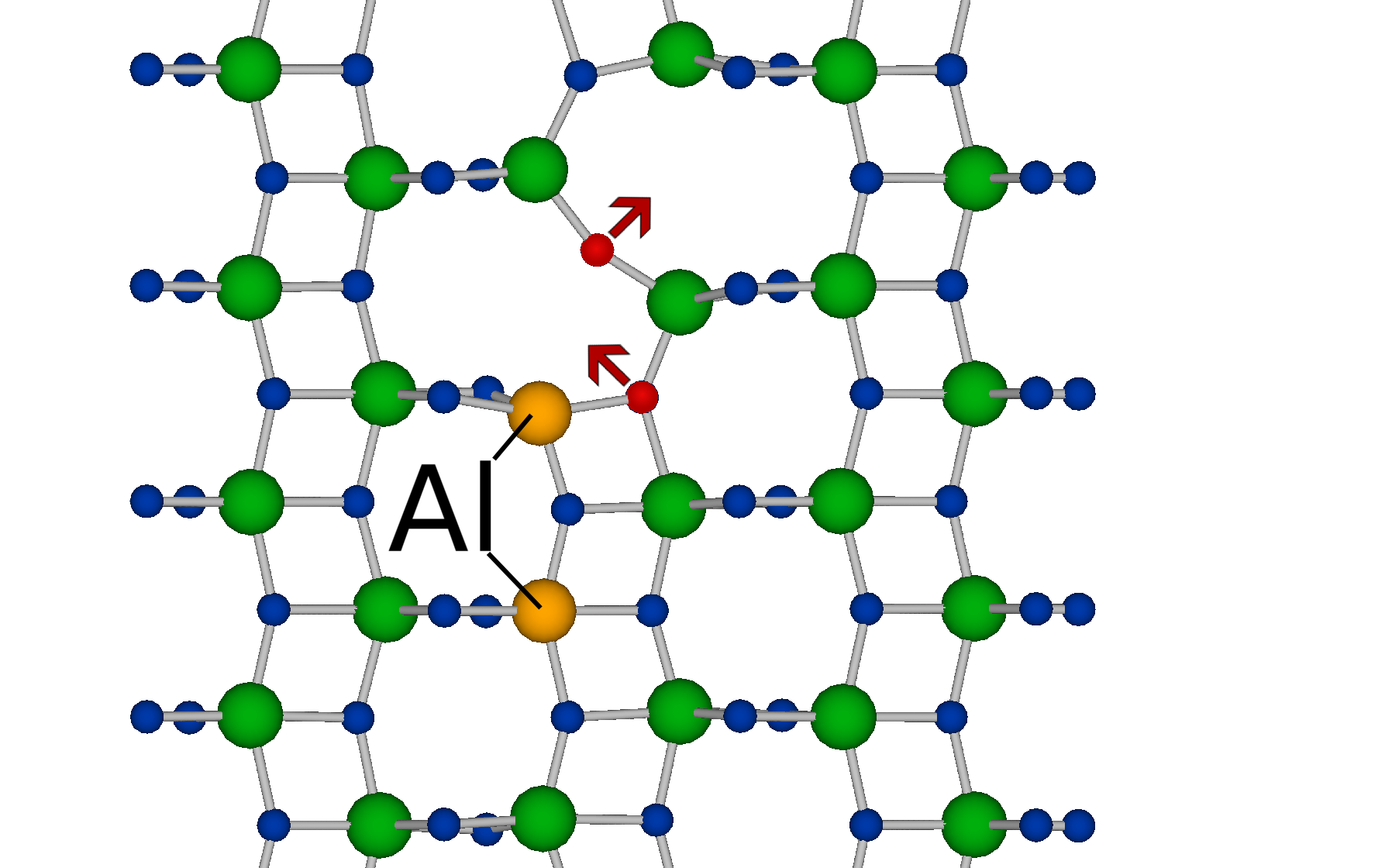}
  \end{minipage}
  &
 \begin{minipage}[t]{0.025\textwidth}
 \vspace{-0.5in}
    \includegraphics[angle=-90,origin=c,trim = 0mm 0mm 0mm 0mm, clip,width=1.2\textwidth]{IMAGES/arrow.pdf}
  \end{minipage}&
  \begin{minipage}[t]{0.31\textwidth}
 \includegraphics[trim = 20mm 0mm 20mm 0mm, clip, width=1.2\textwidth]{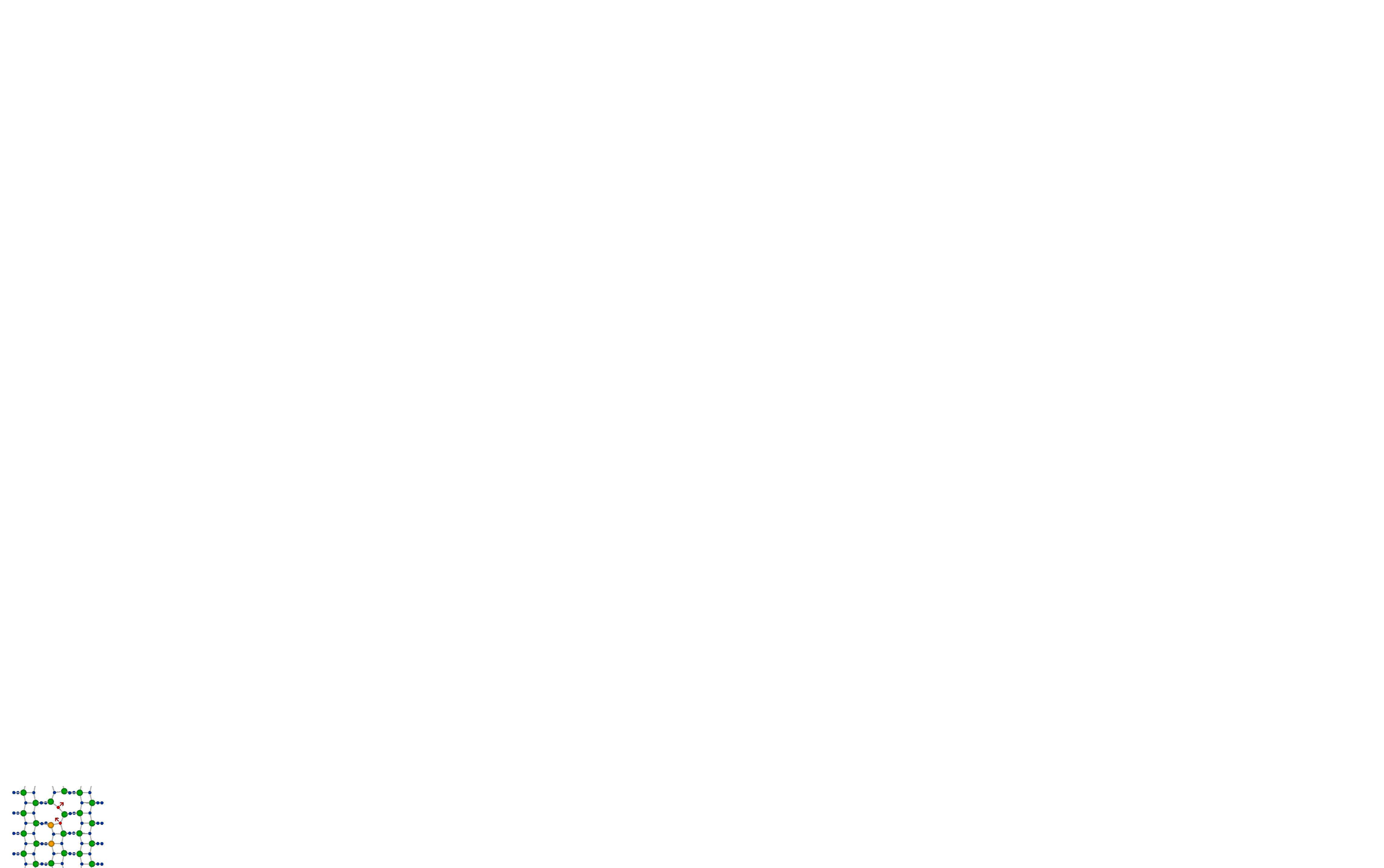}
  \end{minipage}
  &
 \begin{minipage}[t]{0.025\textwidth}
 \vspace{-0.5in}
    \includegraphics[angle=-90,origin=c,trim = 0mm 0mm 0mm 0mm, clip,width=1.2\textwidth]{IMAGES/arrow.pdf}
  \end{minipage}&
  \begin{minipage}[t]{0.31\textwidth}
 \includegraphics[trim = 20mm 0mm 20mm 0mm, clip, width=1.2\textwidth]{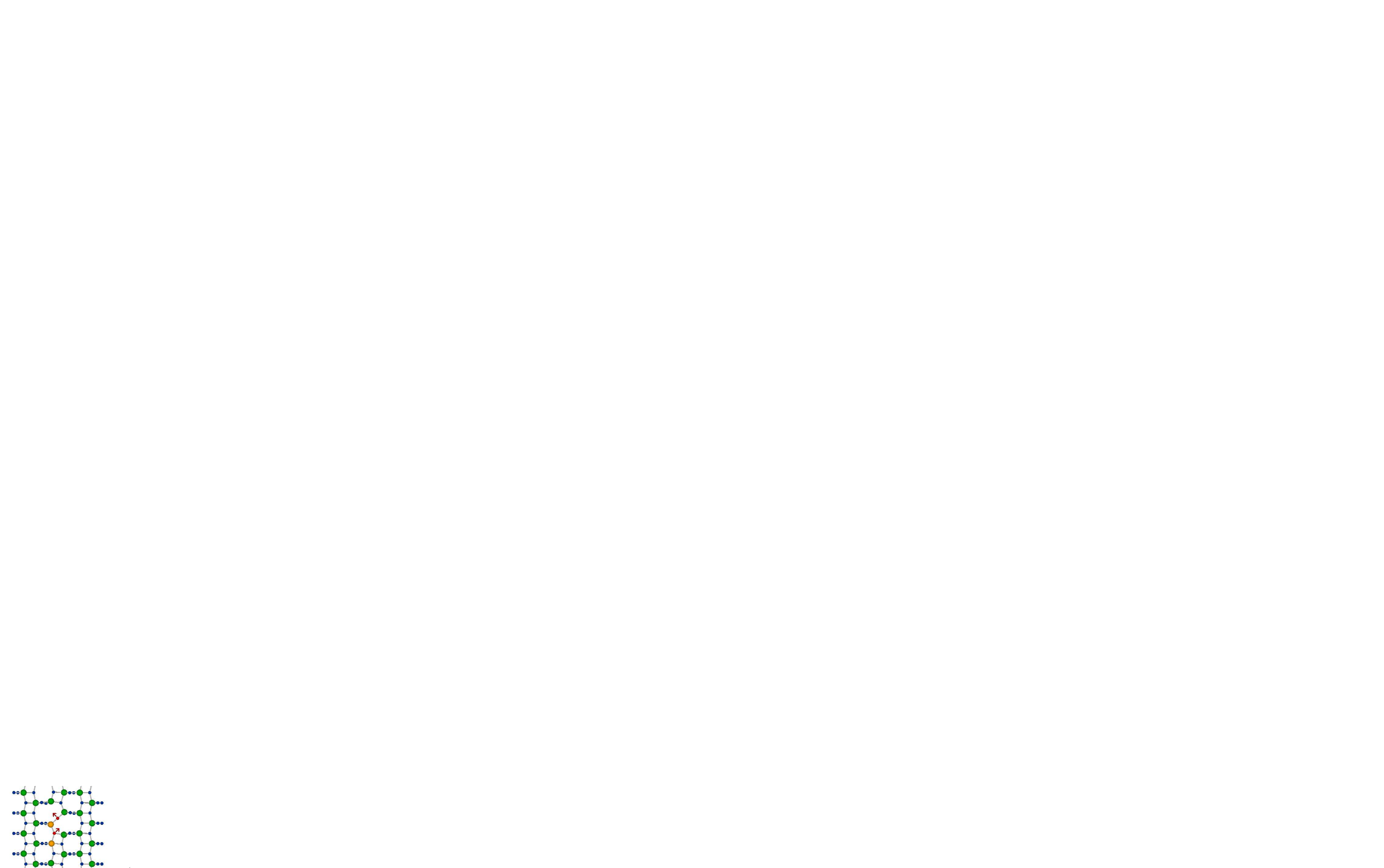}
  \end{minipage}\\
 0&&1&&4\\
&&&&\begin{minipage}[t]{0.025\textwidth}
    \rotatebox[origin=c]{0}{\includegraphics[angle=180,origin=c,trim = 0mm 0mm 0mm 0mm, clip,totalheight=1.2\textwidth]{IMAGES/arrow.pdf}}
  \end{minipage}\\

  \begin{minipage}[t]{0.31\textwidth}
 \includegraphics[trim = 20mm 0mm 20mm 0mm, clip, width=1.2\textwidth]{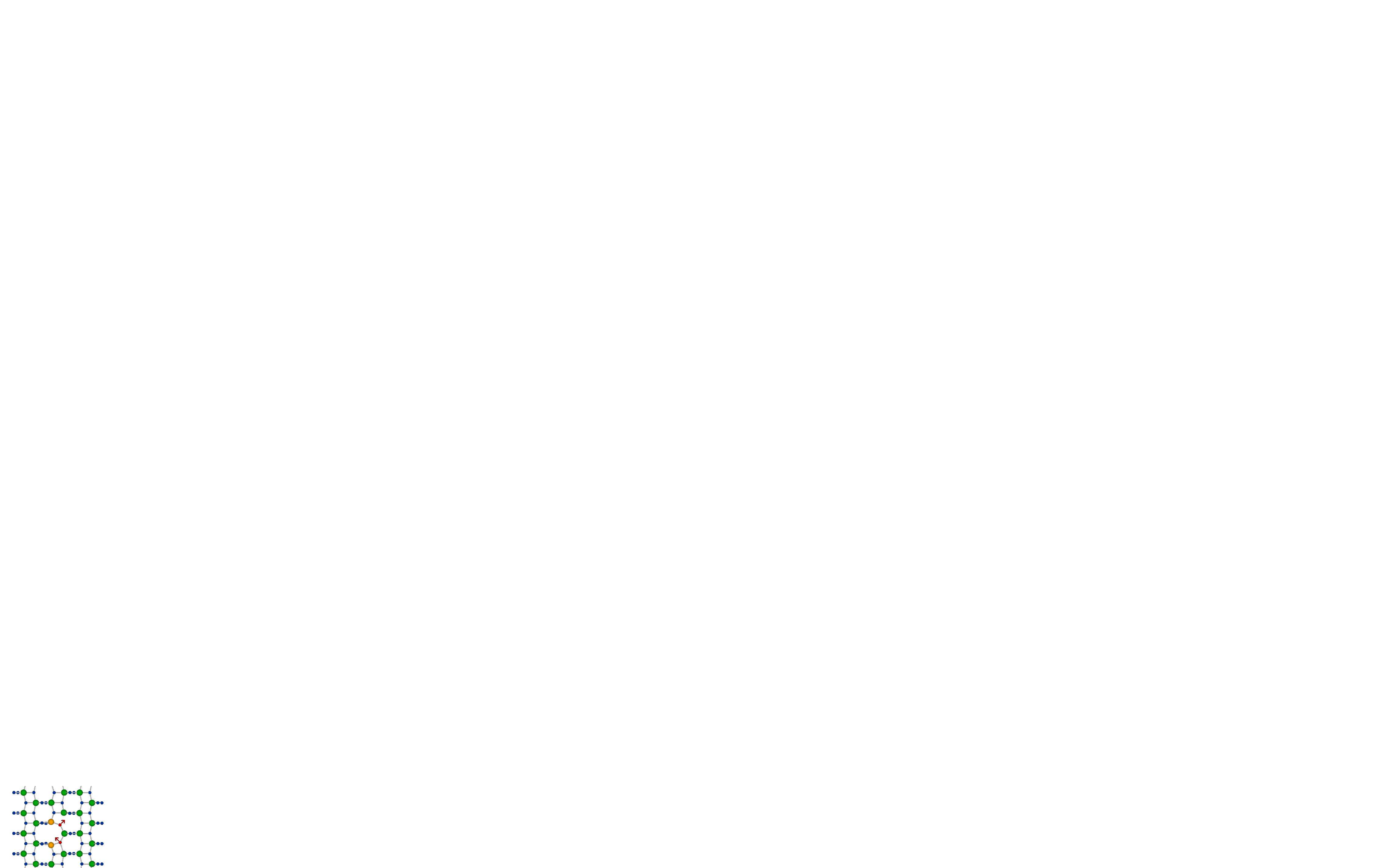}
  \end{minipage}
  &
 \begin{minipage}[t]{0.025\textwidth}
 \vspace{-0.5in}
    \includegraphics[angle=90,origin=c,trim = 0mm 0mm 0mm 0mm, clip,width=1.2\textwidth]{IMAGES/arrow.pdf}
  \end{minipage}&
  \begin{minipage}[t]{0.31\textwidth}
 \includegraphics[trim = 20mm 0mm 20mm 0mm, clip, width=1.2\textwidth]{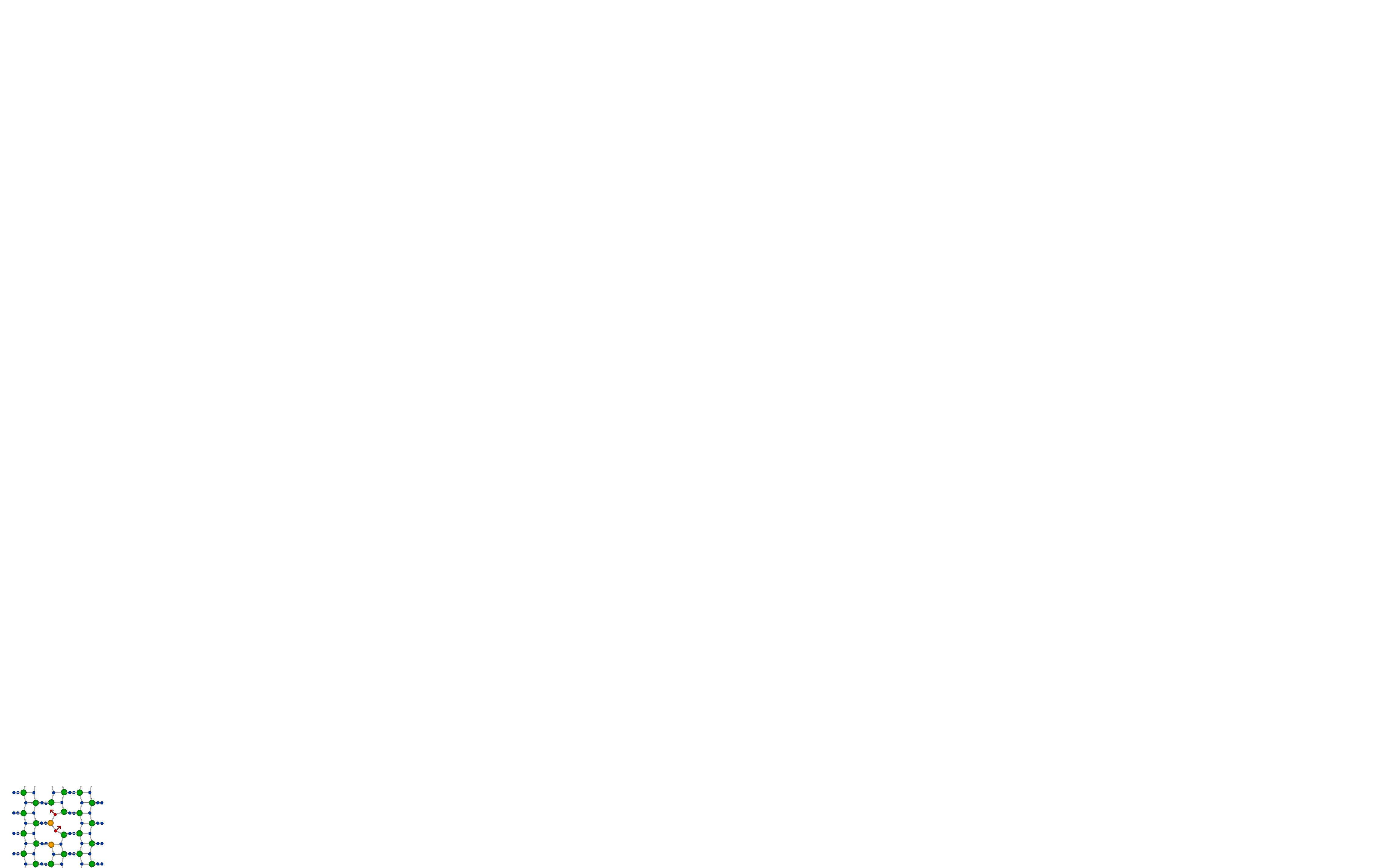}
  \end{minipage}
  &
 \begin{minipage}[t]{0.025\textwidth}
 \vspace{-0.5in}
    \includegraphics[angle=90,origin=c,trim = 0mm 0mm 0mm 0mm, clip,width=1.2\textwidth]{IMAGES/arrow.pdf}
  \end{minipage}&
  \begin{minipage}[t]{0.31\textwidth}
 \includegraphics[trim = 20mm 0mm 20mm 0mm, clip, width=1.2\textwidth]{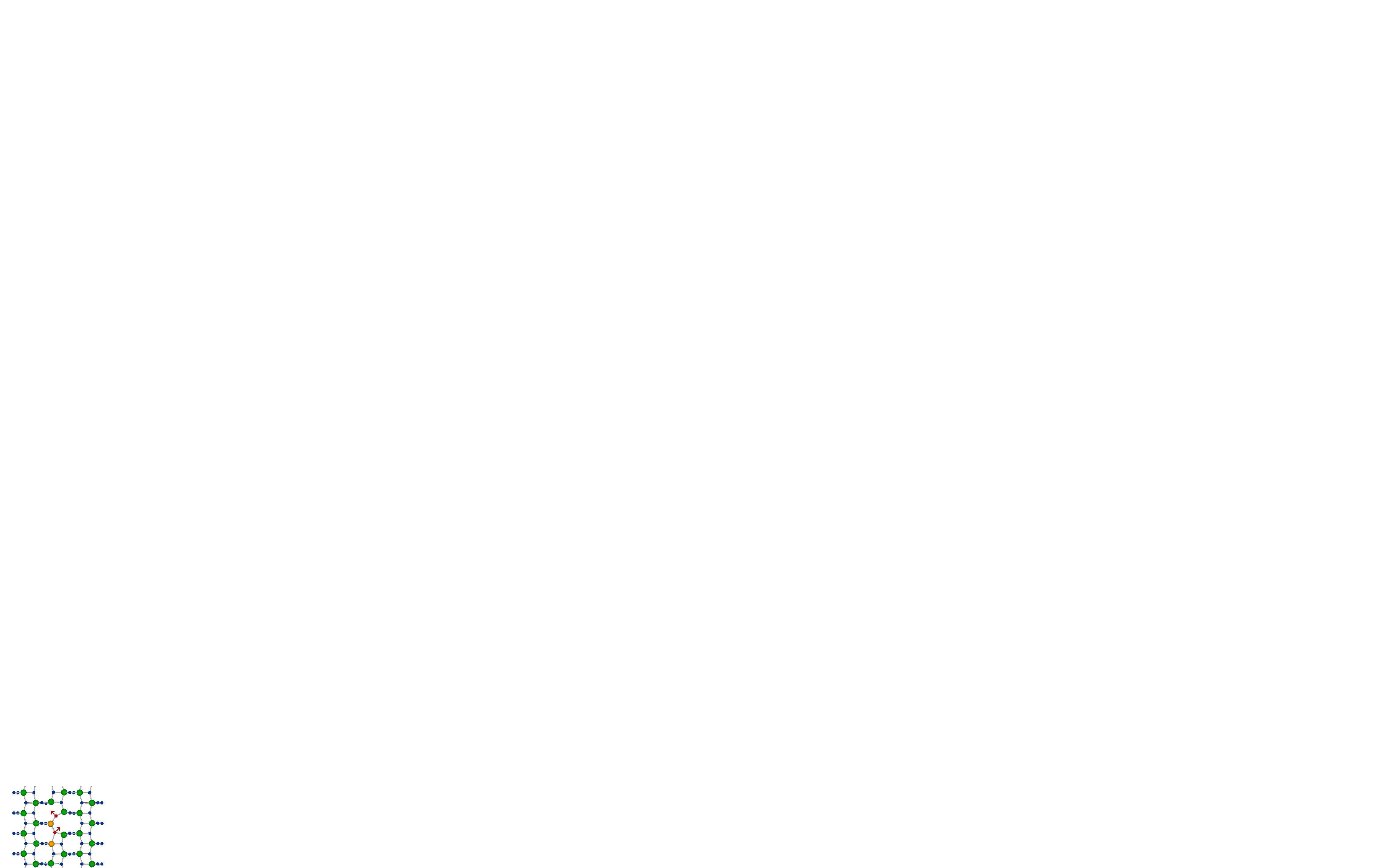}
  \end{minipage}\\
7&&6&&5\\

\end{tabular}
\end{minipage}
&
\hspace{0.05\textwidth}
\begin{minipage}[c]{0.5\textwidth}
\vspace{-0.1\textwidth} \includegraphics[width=1.\textwidth]{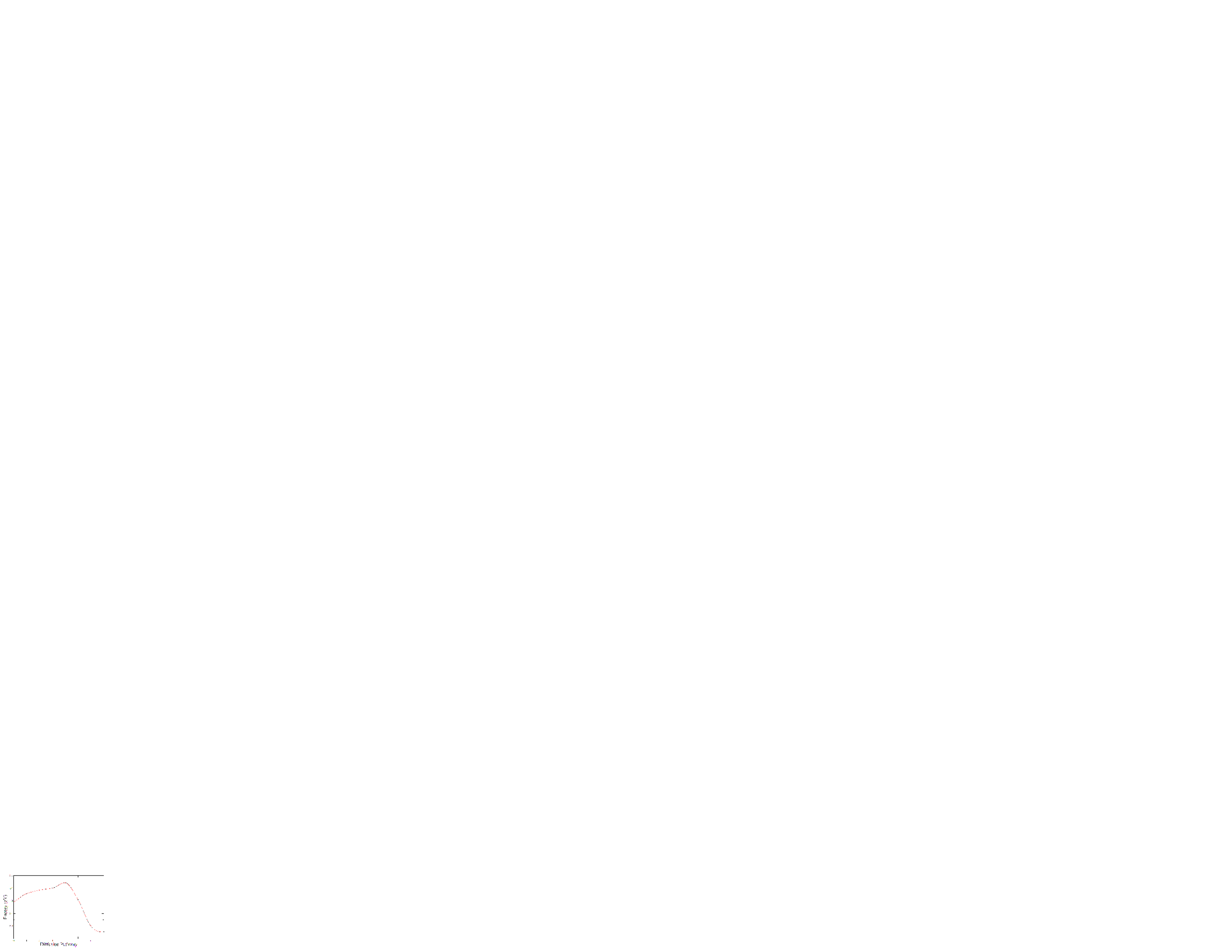}
\end{minipage}\\
\textbf{(a)}&\textbf{(b)}\\
\end{tabular}
\caption{\textbf{(a)} Diffusion pathway for an oxygen vacancy towards Aluminium dopants resulting in a defect of A2 type. For clarity only the central layer is illustrated. \textbf{(b)} Potential energy pathway along the shown diffusion pathway, with a spline fitted to the data to serve as a guide to the eye.}
 \label{A2-diff}

\end{figure}

Next we consider the formation of A2 and A3 defects, again through the diffusion of
oxygen vacancies towards Al dopants.  
It is worth noting that we have performed
short tests using the NEB approach on the likelihood of substitutional 
Al dopants being able to traverse
the lattice mediated by vacancies, but found significant energy barriers suggesting
that Al dopants are indeed immobile.
However given an experimental atomic composition of 3.3\% aluminium in the doped 
TiO\subscript{2} powders\cite{Ko-AL-DSSC}, and assuming that Aluminium is doped 
substitionally, Al dopants will occupy $\sim10\%$ of all lattice sites available.
Each Ti lattice site is "coordinated" to 8 other Ti lattice sites, 
by which we mean there are 8 other adjacent Ti lattice sites 
which, when occupied by aluminium atoms, will arise in the proper 
configuration for either an A2 or A3 defect to be formed. Therefore the 
probability of two Al dopants residing adjacent to one another 
on the lattice of Ti sites in this manner is very significant. For 
example in a bulk supercell
 containing 60 atoms, at the experimental composition 
of 3.3\%, there will be around two dopants. Each Ti lattice position
available to these dopants is coordinated in the manner discussed
to over 40\% of remaining Ti lattice sites. 

Therefore we examine the energetic barrier for oxygen vacancy diffusion
towards two Al atoms residing adjacent to one another in the 
TiO\subscript{2} lattice, in order to gauge the typical energetic barrier

\begin{figure}
\begin{tabular}{c c}
\begin{minipage}[t]{0.45\textwidth}
\hspace{-0.1\textwidth}
\begin{tabular}{c l c l c}
  \begin{minipage}[t]{0.31\textwidth}
 \includegraphics[trim = 20mm 0mm 30mm 0mm, clip, width=1.2\textwidth]{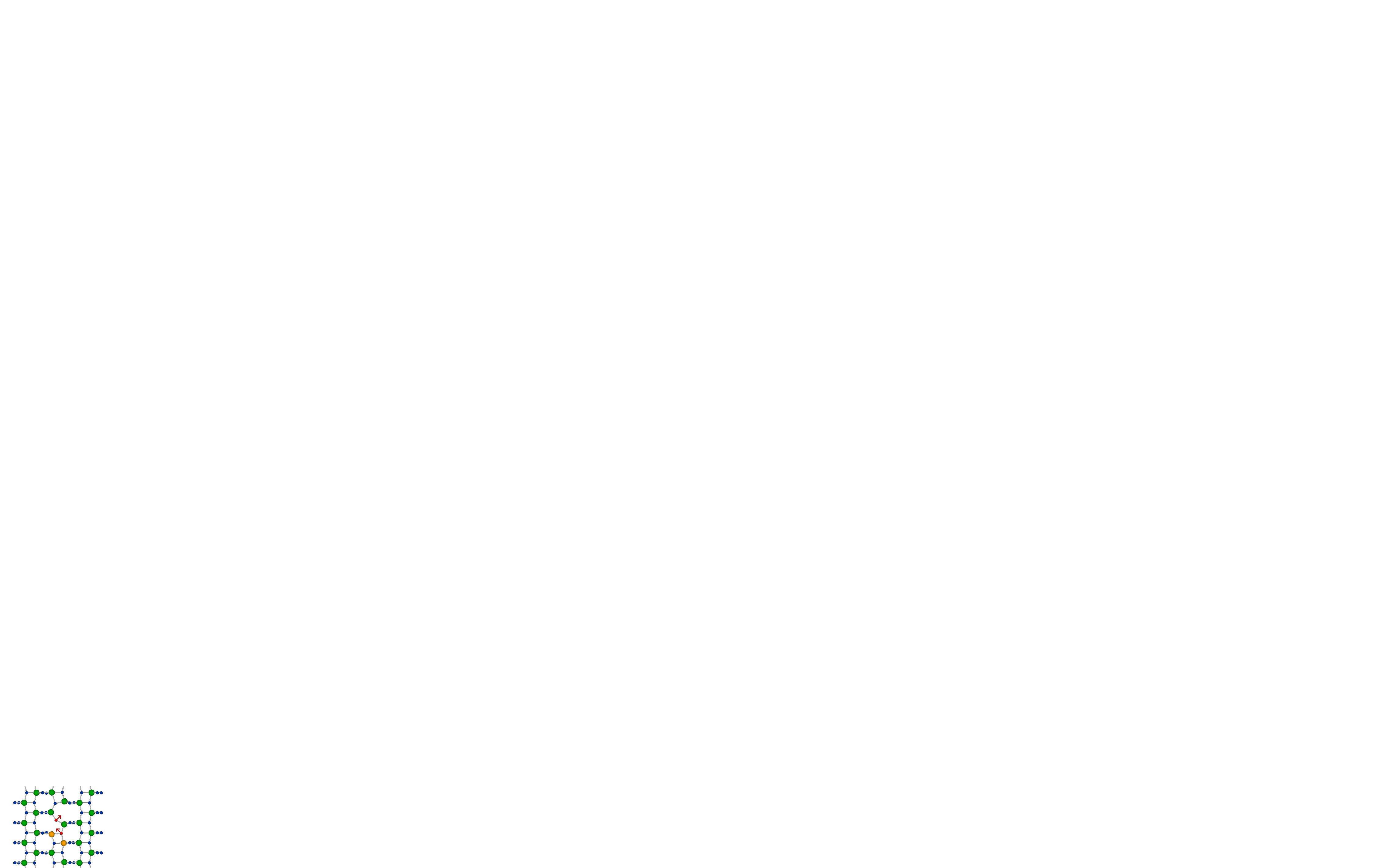}
  \end{minipage}
  &
 \begin{minipage}[t]{0.025\textwidth}
 \vspace{-0.55in}
    \includegraphics[angle=-90,origin=c,trim = 0mm 0mm 0mm 0mm, clip,width=1.2\textwidth]{IMAGES/arrow.pdf}
  \end{minipage}&
  \begin{minipage}[t]{0.31\textwidth}
 \includegraphics[trim = 20mm 0mm 30mm 0mm, clip, width=1.2\textwidth]{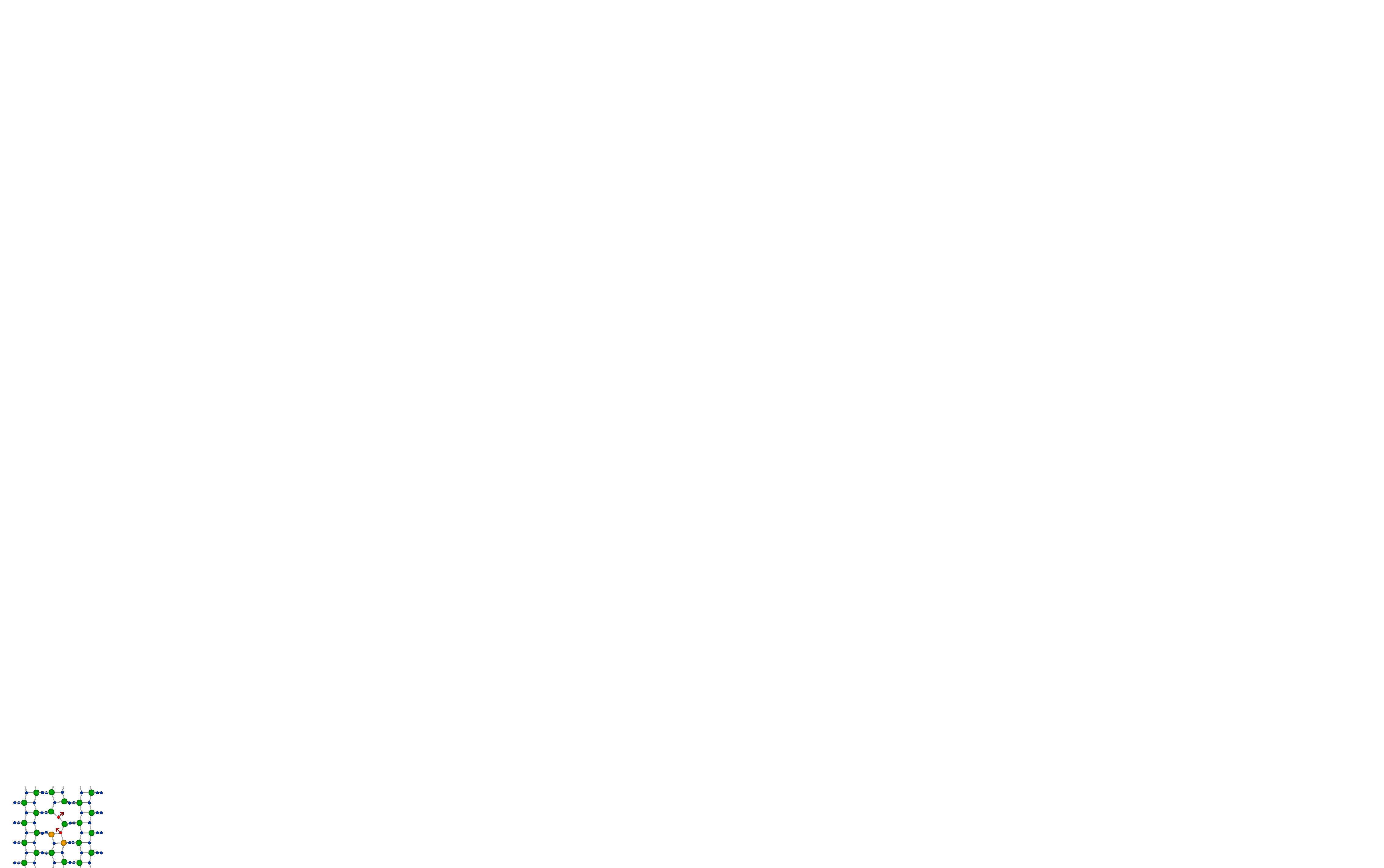}
  \end{minipage}
  &
 \begin{minipage}[t]{0.025\textwidth}
 \vspace{-0.55in}
    \includegraphics[angle=-90,origin=c,trim = 0mm 0mm 0mm 0mm, clip,width=1.2\textwidth]{IMAGES/arrow.pdf}
  \end{minipage}&
  \begin{minipage}[t]{0.31\textwidth}
 \includegraphics[trim = 20mm 0mm 30mm 0mm, clip, width=1.2\textwidth]{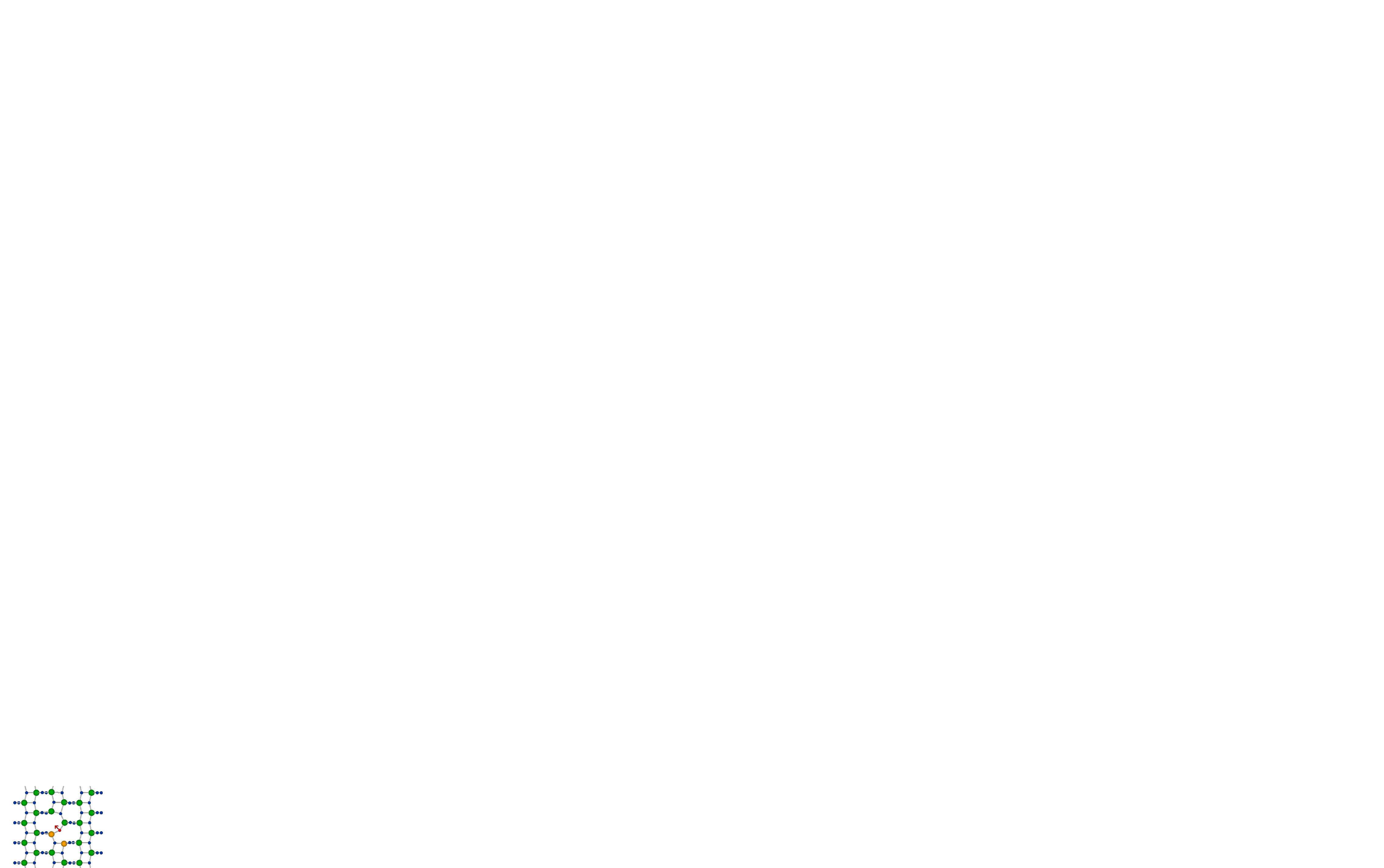}
  \end{minipage}\\
 0&&2&&5\\
&&&&\begin{minipage}[t]{0.025\textwidth}
    \rotatebox[origin=c]{0}{\includegraphics[angle=180,origin=c,trim = 0mm 0mm 0mm 0mm, clip,totalheight=1.2\textwidth]{IMAGES/arrow.pdf}}
  \end{minipage}\\

  \begin{minipage}[t]{0.31\textwidth}
 \includegraphics[trim = 20mm 0mm 30mm 0mm, clip, width=1.2\textwidth]{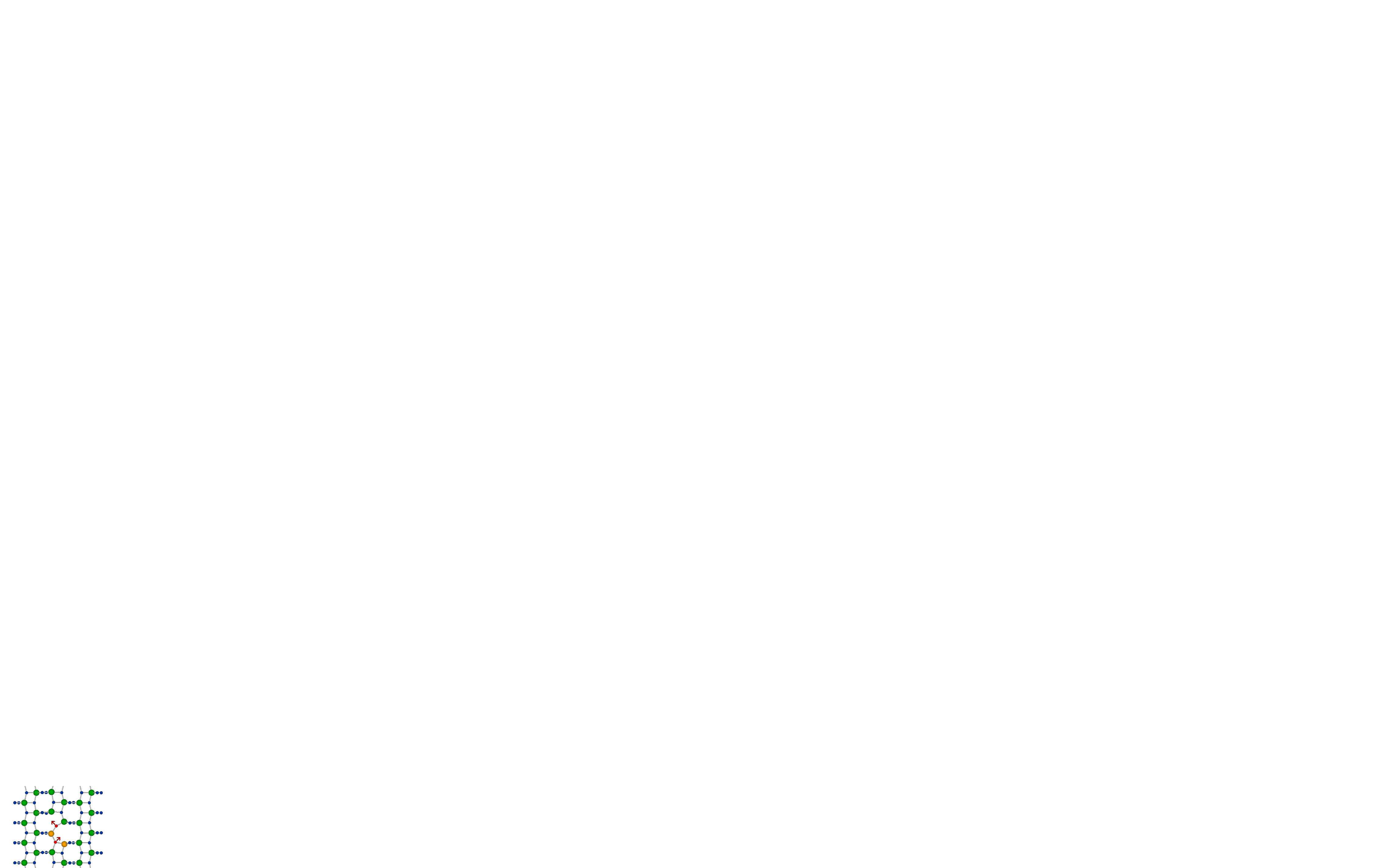}
  \end{minipage}
  &
 \begin{minipage}[t]{0.025\textwidth}
 \vspace{-0.55in}
    \includegraphics[angle=90,origin=c,trim = 0mm 0mm 0mm 0mm, clip,width=1.2\textwidth]{IMAGES/arrow.pdf}
  \end{minipage}&
  \begin{minipage}[t]{0.31\textwidth}
 \includegraphics[trim = 20mm 0mm 30mm 0mm, clip, width=1.2\textwidth]{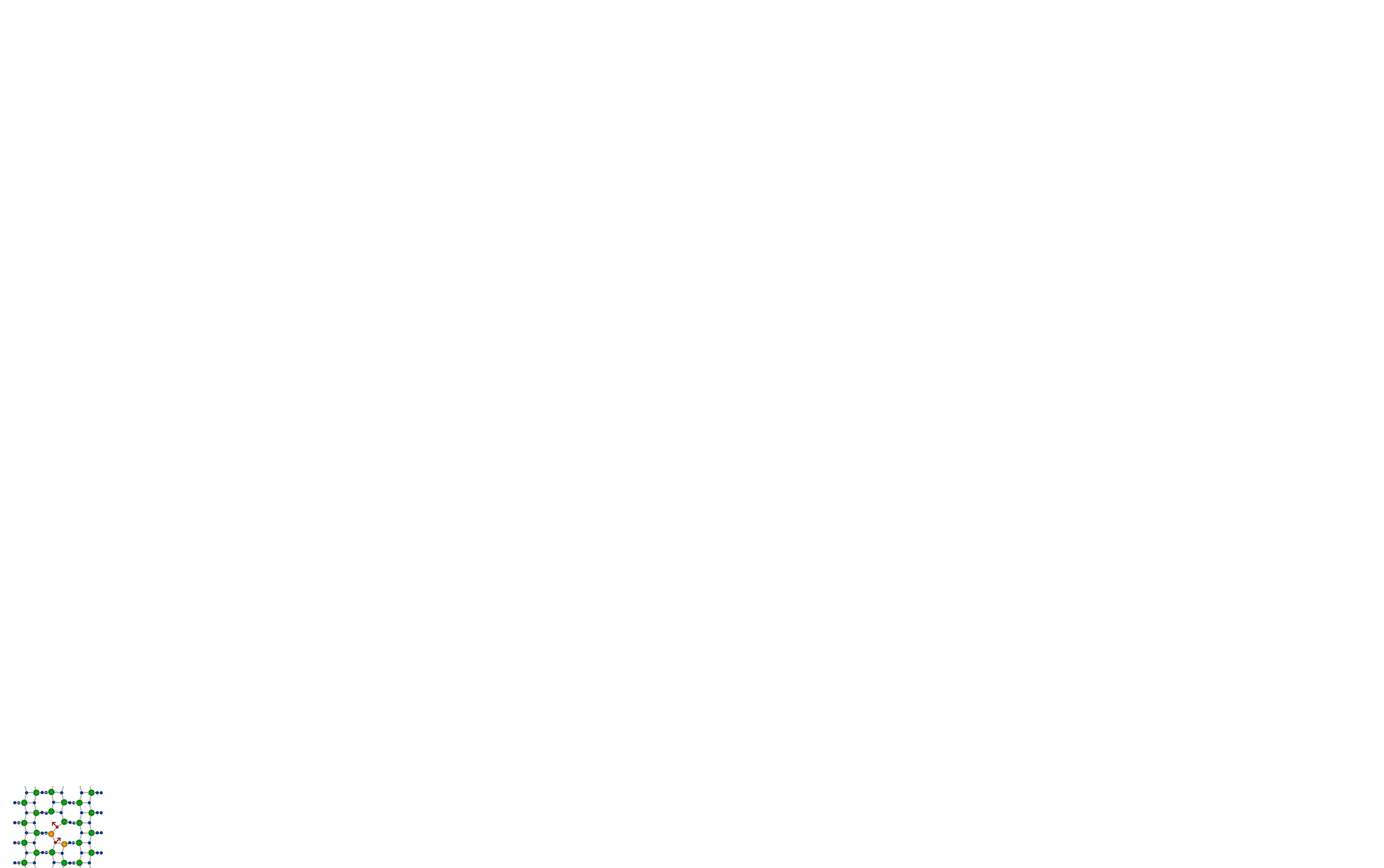}
  \end{minipage}
  &
 \begin{minipage}[t]{0.025\textwidth}
 \vspace{-0.55in}
    \includegraphics[angle=90,origin=c,trim = 0mm 0mm 0mm 0mm, clip,width=1.2\textwidth]{IMAGES/arrow.pdf}
  \end{minipage}&
  \begin{minipage}[t]{0.31\textwidth}
 \includegraphics[trim = 20mm 0mm 30mm 0mm, clip, width=1.2\textwidth]{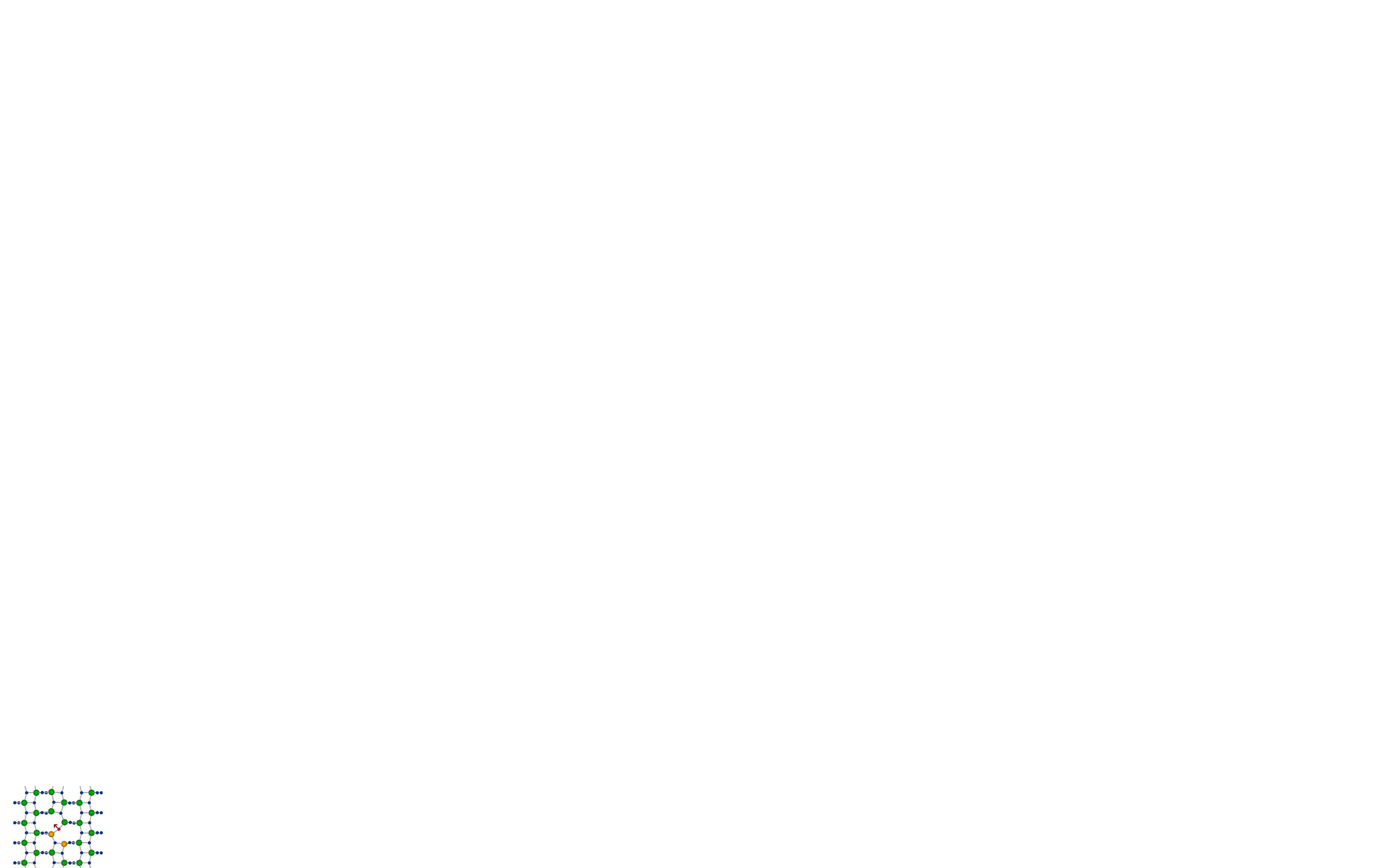}
  \end{minipage}\\
9&&8&&6\\

\end{tabular}
\end{minipage}
&
\hspace{0.05\textwidth}
\begin{minipage}[c]{0.5\textwidth}
\vspace{-0.1\textwidth}\includegraphics[angle=-90,width=1.\textwidth]{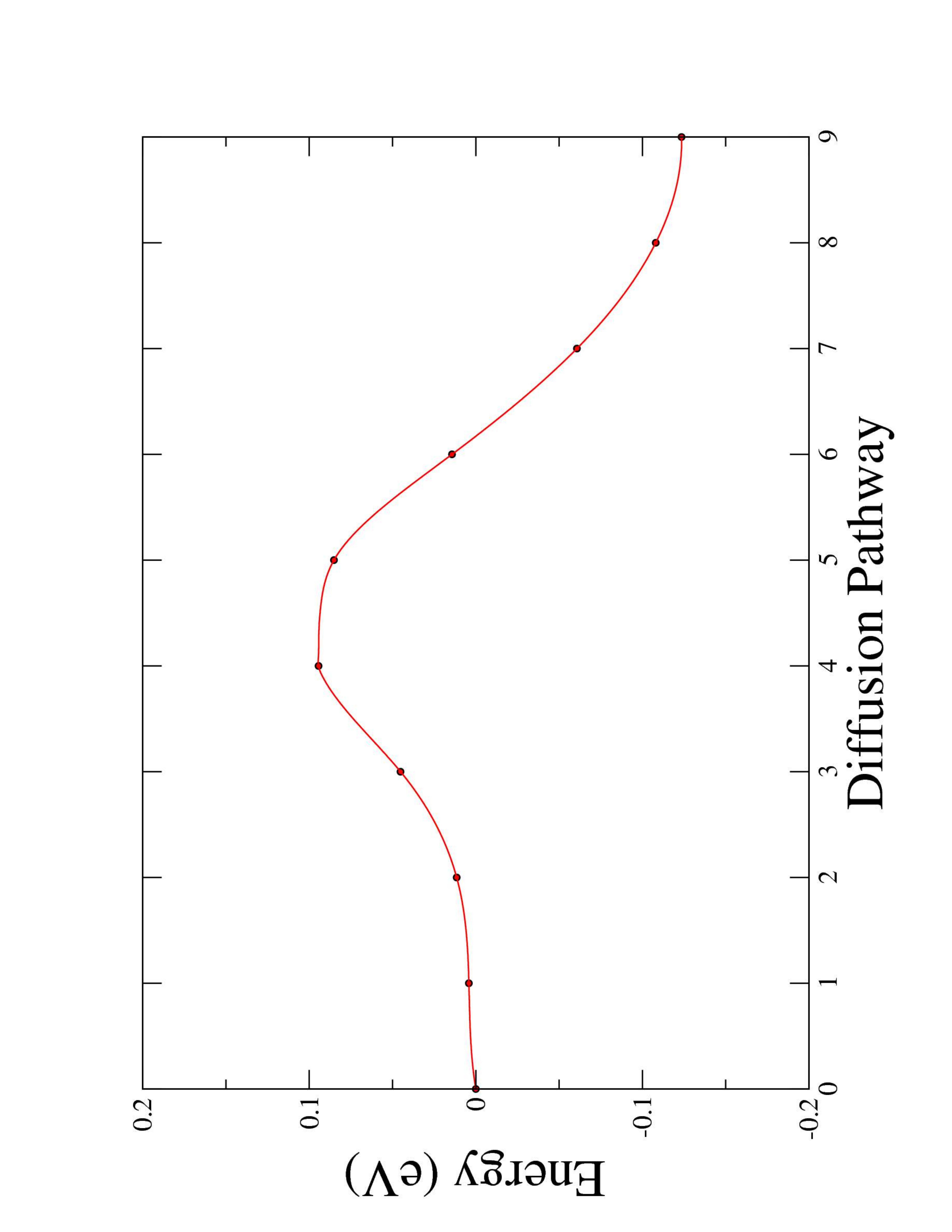}
\end{minipage}\\
\textbf{(a)}&\textbf{(b)}\\
\end{tabular}
\caption{\textbf{(a)} Diffusion pathway for oxygen vacancy towards Aluminium dopants resulting in a defect of A3 type. For clarity only the central layer is illustrated. \textbf{(b)} Potential energy pathway along the shown diffusion pathway, with a spline fitted to the data to serve as a guide to the eye.}
 \label{A3-diff}

\end{figure}

Potential energy surfaces and vacancy diffusion pathways for the formation of
A2 and A3 defects can be seen in Figs. \ref{A2-diff} and \ref{A3-diff}.
Small diffusion barriers of around 0.15 eV and 0.1 eV are found
for the A2 and A3 defects respectively, which will be easily
overcome at room temperature. Energetic gains on overcoming these
barriers of around 0.25 eV and 0.13 eV for A2 and A3 defects illustrate a
bias towards the formation of these defects. Oxygen vacancy diffusion therefore
provides a mechanism by which the aluminium dopants combine with oxygen
vacancies in their proximity. This result explains the experimentally observed
reduction in Ti\superscript{3+} defects on doping of TiO\subscript{2}
 with aluminium \cite{Ko-AL-DSSC}.

\section{Electronic Structure \& Chromophore Adsorption}

\begin{figure}
\begin{center}
   \begin{minipage}[t]{0.65\textwidth}
 \includegraphics[trim = 30mm 0mm 30mm 0mm, clip, width=1.\textwidth]{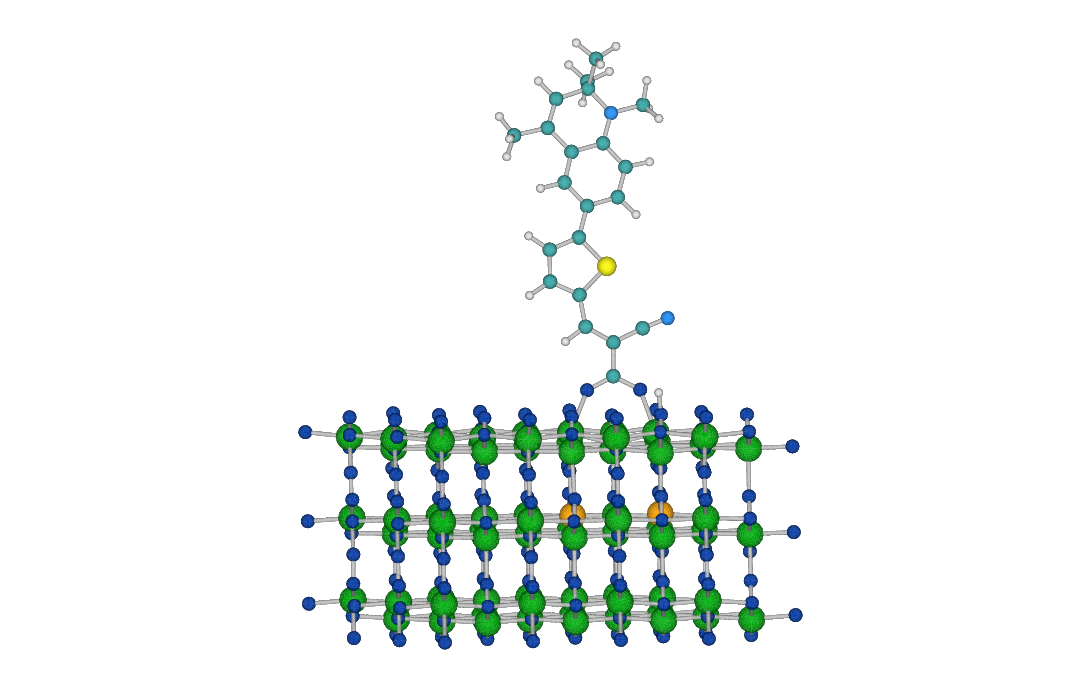}
  \end{minipage}
\caption{C2-1 chromophore adsorbed above a defect of A2 type in the anatase (101) surface.}
 \label{C2-1_pos}
\end{center}
\end{figure}

Adsorption of a chromophore to TiO\subscript{2} electrodes is one of the fundamental interactions in a dye sensitised solar cell. To understand the impact of aluminium doping on this interaction calculations on a composite system in which a chromophore is adsorbed
onto both clean and defective surfaces have been performed. 
For this purpose we have chosen the C2-1 
tetrahydroquinoline chromophore which has shown to successfully
 sensitise TiO\subscript{2} in DSSCs \cite{Tetraq}. The C2-1 chromophore is known 
to preferentially bind to the clean anatase (101) surface 
via a dissociative bidentate bridging mechanism \cite{Conn-Thio} and it is this
adsorption mode that is used throughout. The C2-1 dye has been adsorbed on 
surfaces containing the stable A2 and A3 clustered defects in subsurface
positions, and compared with the adsorption
on a surface containing an oxygen vacancy for comparison. 
Adsorption energies are calculated by the subtraction of the energy of the 
defective slab and the isolated C2-1 chromophore from that of the total system and 
are shown in Table \ref{Ads-En}. Adsorption on the slab containing 
a sub-surface oxygen vacancy increases the adsorption energy significantly with
respect to the clean surface, whereas the adsorption on slabs containing the 
aluminium defects vary only slightly with respect to the clean surface. 
Experimentally it has been reported that aluminium doped TiO\subscript{2} binds 
chromophores more strongly to the surface, due to a preference for dye molecules to
attach to stable Ti\superscript{4+} atoms rather than Ti\superscript{3+}
\cite{Ko-AL-DSSC}. Our result that the C2-1 adsorption energy increases on the
 preferential subsurface 
oxygen vacancy compared to the Al defects suggests that this result is due to 
observed morphology changes resulting from the doping, not as a result of a 
decrease in Ti\superscript{3+} concentration.

\begin{figure}
\begin{center}
   \begin{minipage}[t]{0.65\textwidth}
 \includegraphics[trim = 0mm 0mm 0mm 0mm, clip, width=1.\textwidth]{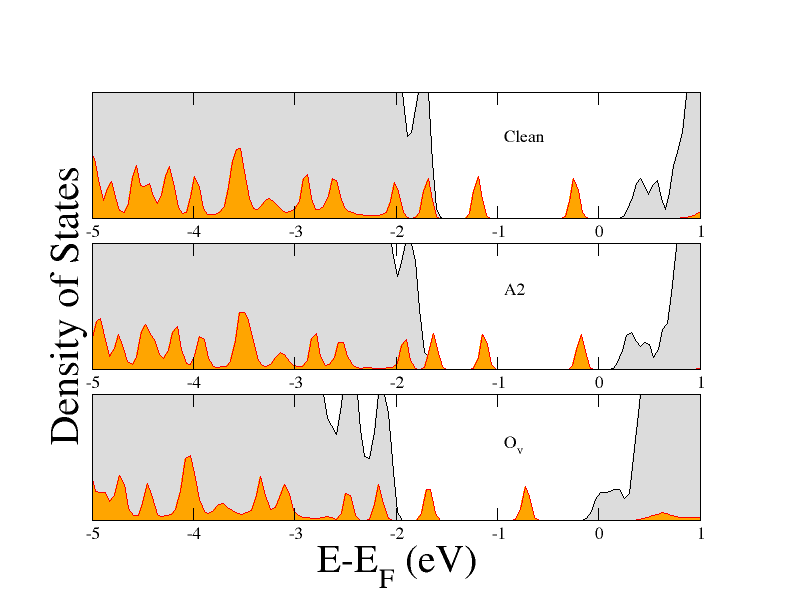}
  \end{minipage}
\caption{Partial density of states. Top: C2-1 adsorbed on clean anatase (101). Middle: C2-1 adsorbed on anatase (101) containing
an A2 subsurface defect. Bottom: C2-1 adsorbed on anatase (101) containing an
oxygen vacancy. The projection on the chromophore is in orange, total DOS in grey.}
 \label{PDOS}
\end{center}
\end{figure}

\begin{table}
\begin{center}
\begin{tabular}{c c c c}
\hline
\hline
Defect & Adsorption energy\\
 type  &       (eV)       \\
\hline
Clean \cite{Conn-Thio}  &  -1.14  \\
O\subscript{V}          &  -1.55  \\
A2                      &  -1.20  \\
A3                      &  -1.12  \\
\hline
\hline
\end{tabular}
\caption{Calculated adsorption energies for the C2-1 chromophore on the anatase (101) surface.}
\label{Ads-En}
\end{center}
\end{table}

As we have already seen that in the A1 substitutional defect
an Al\superscript{3+} ion replacing a Ti\superscript{4+} ion results in a
valence band hole, while an oxygen vacancy introduces occupied Ti\superscript{3+} states
into the band-gap. The combination of two substitutional
aluminium atoms with an oxygen
vacancy (A2 \& A3) causes the formal charge to be maintained,
 with the two
 electrons from the oxygen vacancy filling the valence band
holes resulting from
Al\superscript{3+} substitutions. Oxygen vacancy states
 are not present and the substitutional valence band holes disappear, with the
result that the doped A2 and A3 anatase surfaces behave as if they were clean. We can see this result on examining the density of
states for the C2-1 adsorbed on the A2 defect and on the clean
surface, Fig. \ref{PDOS}, with little difference
between the electronic structure for C2-1 adsorbed on the
clean surface and that of the C2-1 adsorbed on the A2 (a similar result is found for the A3 defect).



Adsorption above the oxygen vacancy        
results in electronic structure that retains the occupied
 defect state at the bottom
of the conduction band. This can be seen in Fig. \ref{PDOS} where the
HOMO (at the zero on the x-axis) resides in the conduction
band for the O\subscript{v} case, while it is dye localised for the 
A2 \& A3 cases (it is worth noting here that the result in Fig. \ref{PDOS} for the oxygen vacancy was obtained including the effects
of spin polarisation but, with little difference observed between 
spin up and spin down, we report only the spin up result).
 
\begin{figure}
\begin{center}
   \begin{minipage}[t]{0.65\textwidth}
 \includegraphics[angle=-90,trim = 60mm 30mm 30mm 60mm, clip, width=1.\textwidth]{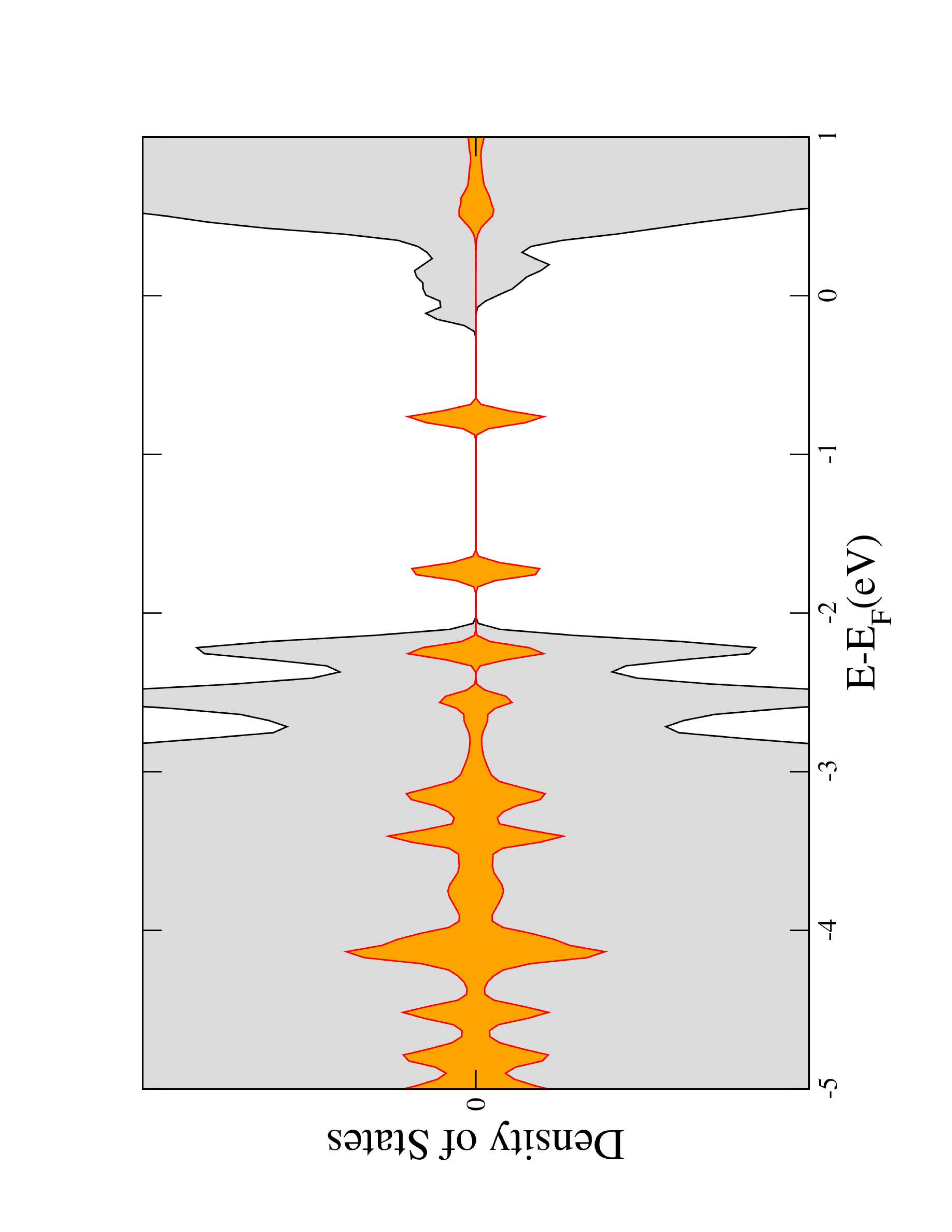}
  \end{minipage}
\caption{GGA+U partial density of states for C2-1 adsorbed above an oxygen vacancy. The projection on the chromophore is in orange, total DOS in grey.}
 \label{pdos_Ovac}
\end{center}
\end{figure}

Interestingly the oxygen vacancy
also down shifts the dye localised HOMO
relative to the conduction band, while the A2 defect very slightly shifts
these dye localised gap states closer to the conduction band. We have also performed
adsorption calculations on the oxygen vacancy with GGA+U (again
using the PAW approach) and found a similar result, as can be seen
in Fig. \ref{pdos_Ovac}, with the oxygen
vacancy defect states being retained just below the conduction band
and a down-shift of the dye HOMO level within the band gap (in this
case application of the GGA+U method provides the spin polarised
result). As well as providing the reported recombination 
Ti\superscript{3+} centres this shifting of the HOMO level by the introduction of an oxygen vacancy could have consequences for DSSC efficiency,
potentially compounding the effect of recombination by adding to
the reduction in J\subscript{SC} and V\subscript{OC}.

\section{Conclusions}

Density functional theory calculations on the Al\superscript{3+} doping of
anatase TiO\subscript{2} have been performed, with both the bulk and (101)
surface examined. Single substitutions of Al\superscript{3+} with Ti\superscript{4+}
 (defect A1) and
clustering of two of these extrinsic Al\superscript{3+} dopants with an intrinsic
 oxygen vacancy have been investigated (defect type A2 and A3).
 Defect types A1 were found to be endothermic in the bulk, and exothermic at 
sub-surface sites on the (101) surface. A2 and A3 were found to be exothermic in the
bulk and also preferentially occupy sub-surface sites on the (101) surface.

Nudged elastic band calculations have illustrated that combination of an
intrinsic vacancy and extrinsic Al\superscript{3+} dopants to form stable
defects is possible and
likely through oxygen vacancy diffusion. Low energy barriers for this diffusion process
with 
an energetic bias towards the formation of A2 and A3 defects
have been found, and vacancy diffusion has been concluded to be a viable
route to these clustered defects.

Susbstitution of Ti\superscript{4+} with A1\superscript{3+} results in a valence band hole.
 Combining two of these substitutions with an O vacancy result in the
formally neutral defects, A2 and A3, in which the typical oxygen vacancy
Ti\superscript{3+} states are not present and the valence band
hole disappears.
This `cleaning up' of
the oxygen vacancies in the TiO\subscript{2} subsurface by benign aluminium
doping results in a reduction in Ti\superscript{3+} states and explains the observed
increase in DSSC efficiency obtained as a result. 
Recombination at these Ti\superscript{3+} sites effectively leads to a 
reduced J\subscript{SC}, the open circuit voltage can also be reduced as a result of
 a down shift in the quasi-Fermi energy. Similary a reduction in 
J\subscript{SC} could occur as a result of the observed vacancy 
induced downshift in dye localised states in the gap.
Removing these defects can improve both J\subscript{SC} and V\subscript{OC}, and thereby 
improve efficiency.

Adsorption of a typical DSSC chromophore, the C2-1 tetraquinoline dye, on the 
defective surface has been investigated.
Adsorption to the oxygen vacancy site is found to be the most energetically favoured,
with the adsorption on the A2 and A3 defects behaving much like adsorption on a
clean surface. The observed V\subscript{OC} increase on aluminium doping of the TiO\subscript{2}
electrodes in DSSCs is concluded to be as a result of
the reduction in Ti\superscript{3+} states, and not due to stronger binding.


\section*{Acknowledgements}
C.O'R. is supported by the MANA-WPI project through a collaboration with
Cambridge University, and D.R.B. was funded by the Royal Society. 
We thank Umberto Terranova and Rami Louca for useful
discussions. This work made 
use of the facilities of HECToR, the UK's national high-performance computing 
service, which is provided by UoE HPCx Ltd at the University of Edinburgh, Cray 
Inc and NAG Ltd, and funded by the Office of Science and Technology through EPSRC's 
High End Computing Programme. Calculations were performed at HECToR through the UKCP 
Consortium which is funded by EPSRC grant EP/F040105. 
The authors acknowledge the use of the UCL Legion High Performance
 Computing Facility, and associated support services, in the completion of this work.

\section{Supporting Information}
Relaxed GGA+U (U=6eV) coordinates for the A1 defect are included in .xyz format. 
This information is available free of charge via the
Internet at http://pubs.acs.org
\bibliography{AL_DOP}

\begin{tocentry}
\begin{centering}
\includegraphics[scale=0.25]{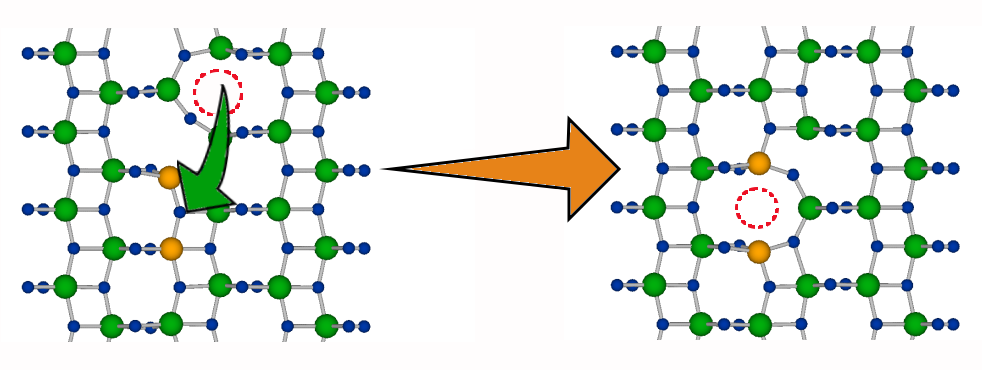}\\

Diffusion of oxygen vacancy in the doped Anatase TiO\subscript{2} (101) sub-surface.
\end{centering}
\end{tocentry}

\end{document}